\documentclass[%
reprint,
amsmath,amssymb,
aps,
floatfix,
superscriptaddress
]{revtex4-2}

\usepackage{graphicx}
\usepackage{bm}
\usepackage[colorlinks=true]{hyperref}
\usepackage{amsfonts}
\usepackage{amsmath}
\usepackage{amssymb}
\usepackage{upgreek}
\usepackage{array} 
\usepackage{multirow}
\begin{document}
	\title{Coherent perfect absorption of anti-modes in an indirect coupled magnon-polariton system}
	\begin{abstract}
		In this work, we report coherent perfect absorption (CPA) of anti-modes in an indirectly coupled magnon--polariton system. By examining both single and indirectly coupled cases, we experimentally distinguish the modal decay rate $\gamma$ from the effective decay rate $\gamma_{\mathrm{eff}}$. At CPA, $\gamma_{\mathrm{eff}} = 0$, leading to a vanishing output and a visually narrow spectrum in the dB-scale, while the intrinsic linewidth set by $2\gamma$ remains unchanged, demonstrating that the effective decay rate dictates the spectral amplitude rather than the physical loss. Furthermore, in the indirectly coupled system, CPA persists over a broad, magnetically tunable detuning range, in contrast to the single-detuning CPA observed in the directly coupled case, thereby enabling magnetically reconfigurable and frequency-selective microwave absorbers.
		
	\end{abstract}
	\author{Chenyang Lu}
	\affiliation{Department of Physics and Astronomy, University of Manitoba, Winnipeg, Manitoba R3T 2N2, Canada}
	\author{Jiguang Yao}
	\affiliation{Department of Physics and Astronomy, University of Manitoba, Winnipeg, Manitoba R3T 2N2, Canada}
	\author{Jiongjie Wang}
	\affiliation{Department of Physics and State Key Laboratory of Surface Physics, Fudan University, Shanghai 200433, China}
	\author{Jiang Xiao}
	\affiliation{Department of Physics and State Key Laboratory of Surface Physics, Fudan University, Shanghai 200433, China}
	\affiliation{Institute for Nanoelectronics Devices and Quantum Computing, Fudan University, Shanghai 200433, China}
	\affiliation{Shanghai Research Center for Quantum Sciences, Shanghai 201315, China}
	\affiliation{Hefei National Laboratory, Hefei 230088, China}
	\author{Can-Ming Hu}\email{hu@physics.umanitoba.ca; \\URL: http://www.physics.umanitoba.ca/$\sim$hu}
	\affiliation{Department of Physics and Astronomy, University of Manitoba, Winnipeg, Manitoba R3T 2N2, Canada}
	\date{\today}
	\maketitle
	
	\section{introduction}  
	
	Non-Hermitian physics provides a unified framework for describing systems that exchange energy, particles, or information with their environment. Such an openness leads to effective Hamiltonians with complex eigenvalues, revealing dynamical and spectral phenomena that lie beyond the scope of conventional Hermitian counterparts \cite{ashida_2020,el2018non}. Research on non-Hermitian physics has consequently attracted a growing interest in photonics \cite{el2018non,miri2019exceptional,wang2023non}, optomechanics \cite{aspelmeyer2014cavity}, solid-state spin ensembles \cite{wu2024third}, and other wave-based fields. In these platforms, systems are commonly characterized by frequency-dependent output responses \cite{pozar2012microwave,haus1984waves}, such as scattering parameters and absorption. In this context, certain responses can be expressed as rational functions of frequency and, therefore, exhibit a well-defined pole–zero structure \cite{pozar2012microwave,haus1984waves,chong2014pt}. The poles encode the system eigenmodes under free oscillation: their real parts specify the mode frequencies, and their imaginary parts define the modal decay rates~$\gamma$ \cite{haus1984waves}. Zeros describe anti-modes (zero scattering modes) \cite{krasnok2019anomalies,sweeney2020rsm,trivedi2024circuit}, with real parts determining the anti-mode frequencies and imaginary parts giving rise to damping-like terms commonly referred to as effective decay rates~$\gamma_{\mathrm{eff}}$ \cite{shen2025polaromechanics,zhang2017exceptional,yang2024anomalous,yang2020unconventional,ballantine2021ptcpa,zhang2023mkn_cpa,zheng2023pta_tis,smith2025_exceptional_antimodes}. 
	
	In addition to their mathematical origin, the two parameters manifest themselves experimentally through distinct spectral signatures. For example, in the total output spectrum, the decay rate $\gamma$ typically sets half of the full width at half maximum (FWHM) \cite{haus1984waves,suh2004nonorthogonal}. By contrast, although termed a “decay,” $\gamma_{\mathrm{eff}}$ is not physically associated with actual decay dynamics; instead, it governs the spectral amplitude and is often interpreted empirically as the linewidth near the extrema \cite{shen2025polaromechanics,zhang2017exceptional}. While $\gamma$ always remains finite due to unavoidable intrinsic and radiative dissipation, $\gamma_{\mathrm{eff}}=0$ can be achieved, resulting in zero output and corresponding to a visually ultra-sharp dip on the dB-scale.

	Based on this perspective, spectral features associated with $\gamma_{\mathrm{eff}}=0$ have been realized across physical settings, including critical coupling \cite{yariv2006optical,yang2024anomalous}, reflectionless scattering states \cite{sweeney2020rsm,qian2023nonhermitian_pzr}, bound states in the continuum \cite{hsu2016bic_review,han2023bound_chiral}, and zero-damping conditions \cite{yang2020unconventional,wang2019nonreciprocity_cmag}. They enable a wide range of applications—from narrow-band isolators \cite{kim2022isolator_cmag,kim2024tw_nra_zr}, enhanced sensing \cite{liu2023phase_BIC, wu2023frequency}, to non-Hermitian mode engineering \cite{Rao2024Braiding,meng2025magnon_polaritons_EP}, and absorption-based signal processing \cite{baranov2017coherent,wang2024high}. Although they are realized under different implementations, these phenomena ultimately originate from interference effects and exhibit transmission or reflection suppression.
	
	Among these mechanisms, coherent perfect absorption (CPA) \cite{chong2010coherent,wan2011time,merkel2015control,romero2016perfect,feng2012coherent,ye2016coherent,lai2024room,cui2024dynamic,bruck2013plasmonic,baldacci2015coherent,vetlugin2019coherent,wang2021coherent,vetlugin2021coherent,roger2015coherent} is unique because it suppresses both transmission and reflection, producing extremely sharp dips in the total output spectrum on the dB-scale. CPA arises from engineered destructive interference between multiple scattering pathways, thereby canceling all outgoing waves at a real frequency corresponding to a zero of the scattering-matrix eigenvalue \cite{chong2010coherent}. Building on single-resonator demonstrations, research has progressed to hybridized systems with direct coupling, arising from the spatial overlap of resonator modes, where additional degrees of freedom, such as detuning and coupling strength, further shape the interference \cite{zhang2017exceptional,shen2025polaromechanics,wang2021coherent}. Under this type of coupling, CPA is often realized at a particular detuning point, i.e., zero detuning. \cite{zhang2017exceptional,wang2021coherent}. Under CPA, The anti-modes simultaneously reach a gain–loss balance and $\gamma_{\mathrm{eff}} = 0$, enabling exploration of PT–symmetry 
	and the corresponding exceptional-point (EP) physics, which has attracted tremendous research interest \cite{zhang2017exceptional,wang2021coherent,ballantine2021ptcpa,zhang2023mkn_cpa,zheng2023pta_tis}.
	
	Despite these advances in directly coupled systems, CPA remains largely unexplored under other coupling mechanisms. In this context, indirect coupling \cite{qian2023nonhermitian_pzr,lu2025temporal_indirect,joseph2025dynamic_longrange_pm,lambert2025coherent_EP_magnon} becomes particularly interesting, as it arises when resonant modes interact through a common waveguide and therefore inherit a traveling wave-mediated hybridization pathway that is fundamentally different from spatial mode overlap. It is therefore of interest to investigate the behavior of the modes and anti-modes in this coupling configuration, as well as their associated $\gamma$ and $\gamma_{\mathrm{eff}}$, and to explore potential new effects enabled by this mechanism, thereby extending CPA studies beyond the well-explored direct coupling regime.
	
	Cavity magnon–polariton (CMP) systems \cite{rameshti2022cavity,wang_2020,harder2021coherent,zhang2014strongly,bai2015spin,harder_2018,yang2020unconventional,tabuchi2014hybridizing,tabuchi2015coherent,lachance2017resolving,li2018magnon,lachance2020entanglement,xu2023quantum}—where magnons hybridize with microwave photons—are among the common platforms for studying indirect coupling \cite{qian2023nonhermitian_pzr,yang2024anomalous, lu2025temporal_indirect}, offering broad tunability through magnetic-field and spatial-mode control, making them naturally suited for this task. Within this platform, extensive studies have reported a variety of phenomena associated with coupling effects, including level repulsion and attraction \cite{artman1955measurement,huebl2013high,goryachev2014high,tabuchi2014hybridizing,bai2015spin,harder_2018,li2019strong,hou2019strong,yang2020unconventional,harder2021coherent,nair2022cavity}, bound states in the continuum \cite{yang2020unconventional,han2023bound_chiral}, and nonreciprocal transport \cite{wang2019nonreciprocity_cmag,yu2020circulating,zhang2020broadband,qian2023nonhermitian_pzr,yaononreciprocity}. CPA has been demonstrated in CMP systems under direct coupling, enabling the realization of exceptional points \cite{zhang2017exceptional} and revealing hybridization across disparate physical modes, such as magnon–polaritons and phonons \cite{shen2025polaromechanics}. In parallel, complete absorption has also been reported in an indirectly coupled CMP system \cite{Rao2024Braiding} under single-port probing, revealing non-Hermitian state braiding with associated topological aspects. However, a systematic understanding of CPA in a conventional two-port indirect coupled setting, particularly from a spectral perspective, still remains lacking. This motivates a detailed examination of such a system, focusing on the mode and anti-mode structure and elucidating the distinct roles of the $\gamma$ and $\gamma_{\mathrm{eff}}$ as manifested in the measured spectra.
	
	In this work, we investigate the decay rate $\gamma$ and the effective decay rate $\gamma_{\mathrm{eff}}$ in both single-resonator and indirectly coupled CMP systems. By examining their pole-zero structure and spectral response, we establish a clear experimental distinction between the two parameters: $\gamma$ controls the FWHM of the total output spectrum, whereas $\gamma_{\mathrm{eff}}$ controls the spectral amplitude. We show that while $\gamma$ remains finite, $\gamma_{\mathrm{eff}}$ can be tuned to zero, leading to the realization of CPA on both single and indirectly coupled CMP platforms. In the indirectly coupled case, CPA persists over a broad detuning range that is tunable via a magnetic field—a behavior not available in directly coupled systems. This unique feature positions CMP systems with indirect coupling as a promising platform for implementing magnetically reconfigurable absorbers.
	
	\section{Theory}
	CPA results from the combined effects of the resonator dynamics and the input-wave configuration, which together produce a scattering zero and eliminate all outgoing waves \cite{chong2010coherent,chong2014pt,wan2011time}. Temporal coupled-mode theory (TCMT)~\cite{fan2003temporal,khan1999mode,zhao2019connection,suh2004nonorthogonal} provides a compact framework to describe this process. Here, we consider the common case where the resonators couple equally to the forward and backward-propagating waves, allowing for simple analytic expressions of the scattering response for single and coupled resonators.
	
	\begin{figure*}
		\centering
		\includegraphics[width=\linewidth]{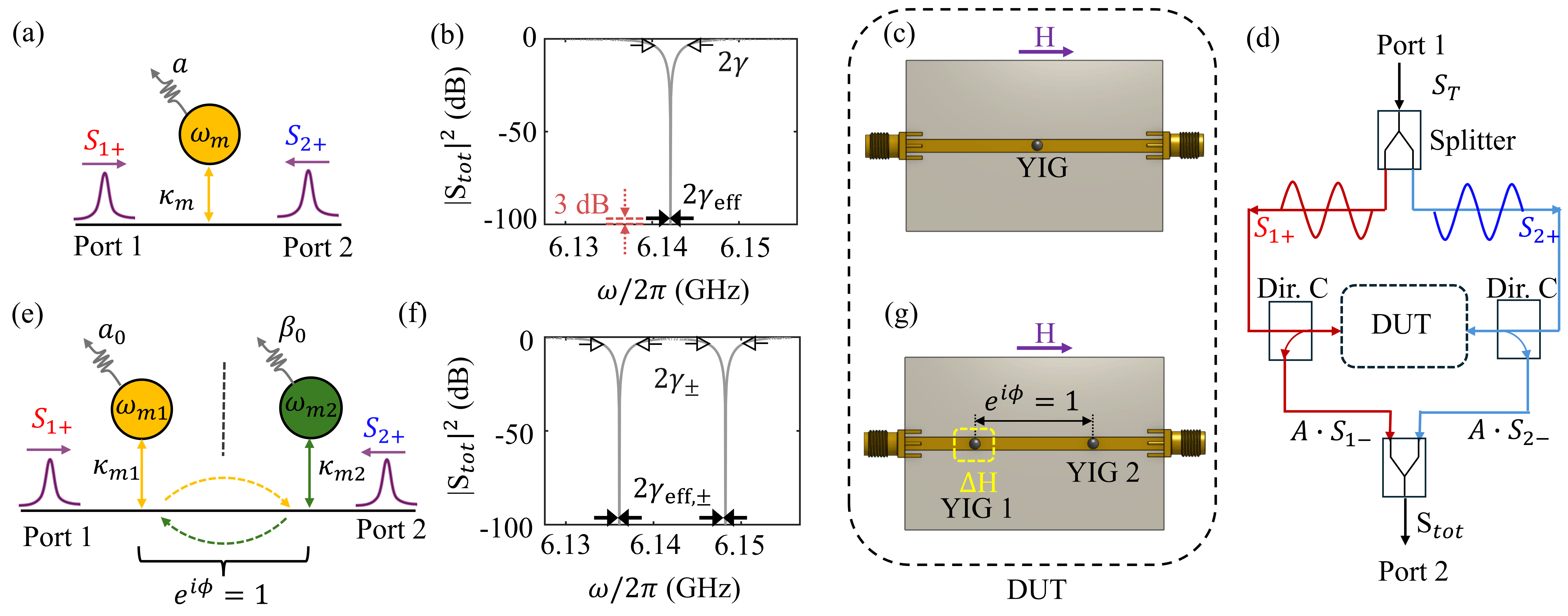}
		\caption{(a),(e) Schematics of the single-resonator and traveling wave-mediated indirectly coupled two-resonator systems under two identical coherent inputs. (b),(f) Calculated total output spectra $|S_{\mathrm{tot}}|^{2}$ for the single and indirectly coupled systems under CPA conditions, based on Eqs.~(\ref{E5}) and (\ref{E14}), respectively. The vanishing $\gamma_{\mathrm{eff}}$ is indicated by the solid arrows (taken from 3 dB above the minimum), while the FWHM indicates the background response is marked by the open arrows. (c) Single-resonator CPA device consisting of a YIG sphere on a transmission line. The bias field $H$ tunes the Kittel-mode frequency. (d) The signal from port 1 of the VNA are split by a symmetric power splitting network to generate two coherent counter-propagating inputs that excite the resonator(s) in the DUT. The outgoing waves are extracted using directional couplers and received at port 2. (g) Indirectly coupled device with two YIG spheres. A global field $H$ is applied to both spheres, while a local tuning field $\Delta H$ is applied to YIG 1 to control the frequency detuning between the two magnon Kittel-modes.}
		\label{Fig1}
	\end{figure*}
	
	\subsection{Single resonator}
	For a single resonator side-coupled to a waveguide under two-port coherent
	excitation, as shown in Fig.~\ref{Fig1}(a), the total outgoing power normalized
	to the total incident power is (see Appendix~\ref{single})
	\begin{equation}\label{E1}
		\begin{split}
			|S_{\mathrm{tot}}|^{2}
			&=
			\frac{
				(\omega-\omega_{m})^{2}
				+\alpha^{2}
				+\kappa_{m}^{2}
				-\dfrac{4 p \alpha \kappa_{m}}{1+p^{2}}
			}{
				(\omega-\omega_{m})^{2}
				+(\alpha+\kappa_{m})^{2}
			}
			\\[4pt]
			&=
			\left|
			\frac{\omega-\tilde{\omega}_{z}}
			{\omega-\tilde{\omega}_{p}}
			\right|^{2}.
		\end{split}
	\end{equation}
	
	Here, $\omega$ is the probe frequency, $\omega_{m}$ is the resonance frequency of
	the resonator, $\alpha$ is the intrinsic loss rate, and $\kappa_{m}$ is the external
	coupling rate to the traveling waves in the waveguide. The parameter $p$ denotes the amplitude ratio between the two input fields $S_{1+}$ and $S_{2+}$, such that $|S_{2+}| = p|S_{1+}|$.
	
	The corresponding complex pole and zero of $S_{\mathrm{tot}}$ are
	$\tilde{\omega}_{p}=\omega_{m}-i\gamma$ and
	$\tilde{\omega}_{z}=\omega_{m}-i\gamma_{\mathrm{eff}}$, whose imaginary parts
	define the decay rate
	\begin{equation}\label{E2}
		\gamma=\alpha+\kappa_{m},
	\end{equation}
	and the effective decay rate;
	\begin{equation}\label{E3}
		\gamma_{\mathrm{eff}}
		=
		\sqrt{
			\alpha^{2}
			+\kappa_{m}^{2}
			-\frac{4p\alpha\kappa_{m}}{1+p^{2}}
		}.
	\end{equation}
	
	Eq.~(\ref{E1}) exhibits a Lorentzian dip, with its FWHM given by the imaginary
	part of the pole, which is $2\gamma$. Consequently, due to the pole-zero structure, $1/|S_{\mathrm{tot}}|^{2}$ can also be written as a Lorentzian lineshape, whose FWHM is given by the imaginary part of the zero, which is $2\gamma_{\mathrm{eff}}$ \cite{yang2020unconventional}.
	
	In addition, we introduce the absorption of the system as a complementary diagnostic \cite{wang_2024,lu2025temporal_indirect}, representing the fraction of input power absorbed by the system. It is defined as
	\begin{equation}\label{E4}
		\mathrm{Abs} = 1 - |S_{\mathrm{tot}}|^{2}.
	\end{equation}

	At $p=0$, indicating input from only port 1, Eq.~(\ref{E1}) reduces to
	\begin{equation}\label{E6}
		\begin{split}
			|S_{\mathrm{tot}}|^{2}
			&=
			\frac{
				(\omega-\omega_{m})^{2}
				+\alpha^{2}+\kappa_{m}^2
			}{
				(\omega-\omega_{m})^{2}
				+(\alpha+\kappa_{m})^{2}
			}
			\\[4pt]
			&=
			|S_{21}|^{2}+|S_{11}|^{2},
		\end{split}
	\end{equation}
	where $S_{21}=(\omega-\omega_{m}+i\alpha)/(\omega-\omega_{m}+i\alpha+i\kappa_{m})$ and $S_{11}=-i\kappa_{m}/(\omega-\omega_{m}+i\alpha+i\kappa_{m})$ are the well-known transmission and reflection coefficients of the single-resonator side-coupled waveguide system, with a detailed derivation provided in Ref.~\cite{khan1999mode}. At $\omega=\omega_{m}$, $|S_{\mathrm{tot}}|^{2}=(\alpha^{2}+\kappa_{m}^2)/(\alpha+\kappa_{m})^{2}$. Under condition $\alpha=\kappa_{m}$, the minimum of $|S_{\mathrm{tot}}|^{2}=0.5$, with $\gamma=4\alpha$ and $\gamma_{\mathrm{eff}}=2\sqrt{2}\,\alpha$ remaining finite.
	
	A particularly interesting case occurs at $p=1$, indicating that the inputs
	from both ports have identical amplitudes. Eq.~(\ref{E1}) reduces to
	\begin{equation}\label{E5}
		\begin{split}
			|S_{\mathrm{tot}}|^{2}
			&=
			\frac{
				(\omega-\omega_{m})^{2}+(\alpha-\kappa_{m})^2
			}{
				(\omega-\omega_{m})^{2}
				+(\alpha+\kappa_{m})^{2}
			}.
		\end{split}
	\end{equation}
	
	In this case, when probing on resonance ($\omega=\omega_{m}$),
	$|S_{\mathrm{tot}}|^{2}=(\alpha-\kappa_{m})^{2}/(\alpha+\kappa_{m})^{2}$. Under
	the condition $\alpha=\kappa_{m}$, $|S_{\mathrm{tot}}|^{2}=0$, corresponding to
	$\mathrm{Abs}=1$ and indicating coherent perfect absorption. Simultaneously,
	the effective decay rate vanishes, $\gamma_{\mathrm{eff}}=0$, while the total
	decay rate remains finite and unchanged, $\gamma=4\alpha$. 
	
	This result can equivalently be derived from the CPA eigenvalue condition of the S-matrix \cite{chong2010coherent}. The CPA condition corresponds to the special case where the S-matrix possesses an eigenvalue $\lambda_{cpa}$ that can reach zero, i.e., $\det(\mathrm{S})=0$, guaranteeing the existence of a specific eigenvector $\mathbf{v}_{\mathrm{CPA}}\neq 0$ satisfying $S\mathbf{v}_{\mathrm{CPA}}=\lambda_{cpa}\mathbf{v}_{\mathrm{CPA}}=\mathbf{0}$. Physically, CPA occurs only when the system is driven under such an eigenvector configuration. 
	
	In the proposed system, the S-matrix has two eigenvalues: one is $-1$ with eigenvector $(1, -1)^T$, corresponding to an anti-absorption state in which all input power is reflected with no absorption. The other eigenvalue, $\lambda_{cpa}$, can reach zero with eigenvector $\mathbf{v}_{\mathrm{CPA}}=(1, 1)^T$. This requires the physical inputs to satisfy 
		$\mathbf{v}_{\mathrm{in}}=(S_{1+}, S_{2+})^T=S_{1+}\mathbf{v}_{\mathrm{CPA}}=S_{1+}(1, 1)^T$, i.e., $S_{1+}=S_{2+}$, meaning equal phase and amplitude of the two inputs, as previous realized for $p=1$. Consequently, $|S_{\mathrm{tot}}|^{2}=|S\mathbf{v}_{\mathrm{in}}|^{2}/|\mathbf{v}_{\mathrm{in}}|^{2}=|\lambda_{cpa}|^2=|{\rm det}(\mathrm{S})|^2$.
	
	Therefore, when the system is driven under the $\mathbf{v}_{\mathrm{CPA}}$ configuration, the zero eigenvalue $\lambda_{cpa}$ and consequently ${\rm det}(\mathrm{S})$ are directly reflected in the total output. This equivalence breaks down for non-$\mathbf{v}_{\mathrm{CPA}}$ excitation, as observed for $p=0$ case.
	
	Fig.~\ref{Fig1}(b) shows a representative spectrum calculated from
	Eq.~(\ref{E5}) on the dB-scale using $\omega_{m}/2\pi=6.14~\mathrm{GHz}$ and
	$\alpha/2\pi=\kappa_{m}/2\pi=0.8~\mathrm{MHz}$. In this spectrum, the decay rate
	$\gamma$ (open arrows) determines the FWHM and thus the broad spectral
	background. 
	
	Although the inverse spectrum $1/|S_{\mathrm{tot}}|^{2}$ provides a rigorous definition of $\gamma_{\mathrm{eff}}$ through its FWHM, taking the inverse essentially acts as a spectral magnifier: as $|S_{\mathrm{tot}}|^{2}$ approaches zero for $\gamma_{\mathrm{eff}} \to 0$, the inverse response diverges. Therefore, $\gamma_{\mathrm{eff}} = 0$ serves as a signature of a vanishing $|S_{\mathrm{tot}}|^{2}$ output amplitude, rather than of a physical linewidth or dissipation.
	
	From this perspective, plotting $|S_{\mathrm{tot}}|^{2}$ on a dB scale plays a similar role: the logarithmic transformation amplifies weak residual signals near the resonance dip, especially close to zero output. As $|S_{\mathrm{tot}}|^{2} \to 0$, the dB response diverges to negative infinity and the apparent dip width collapses, consistent with $\gamma_{\mathrm{eff}} \to 0$. We therefore introduce a phenomenological effective linewidth defined by a 3~dB-above-minimum criterion, which allows for a simple visual extraction of $\gamma_{\mathrm{eff}}$. As an illustrative example, Fig.~\ref{Fig1}(b) shows that the 3~dB-above-minimum linewidth of the narrow resonant dip (solid arrows) collapses to zero under the CPA condition, consistent with $\gamma_{\mathrm{eff}} = 0$.
	
	As the phenomenological 3~dB-above-minimum linewidth and the rigorous inverse-spectrum definition are strictly identical at $\gamma_{\mathrm{eff}} = 0$, they remain within a $\pm5\%$ difference in the vicinity of zero. A detailed quantitative comparison between the two methods is provided in Appendix~\ref{comparison}. 
	
	\subsection{Indirectly coupled resonator}
	Placing two resonators close to a common waveguide induces an indirect coupling between them \cite{yang2024anomalous,qian2023nonhermitian_pzr,lu2025temporal_indirect,wang_2024}, as illustrated in Fig.~\ref{Fig1}(e). Although asymmetric input configurations can lead to distinct and interesting absorption behaviors in a single-oscillator system, the inclusion of an additional mode substantially increases the complexity of the parameter space. To focus on the key role of indirect coupling, we consider only symmetric inputs with equal amplitudes in the coupled system. Additionally, the physical separation between the resonators introduces a distance-dependent traveling phase $\phi$ \cite{pozar2012microwave,lu2025temporal_indirect}. Such a traveling phase controls the waveguide-mediated interference and can induce both coherent and dissipative coupling. Here, the traveling phase $\phi$ is tuned to be $\phi = 2n\pi$ $(n = 0,1,2,\ldots)$ at the operating frequencies, such that the propagation factor satisfies $e^{i\phi}=1$. Consequently, the system operates in the dissipative coupling regime, and the counter-propagating inputs reach both resonators with identical phase. The total outgoing power, normalized to the total incident power, is (see Appendix~\ref{double})
	\vspace{1pt}
	\begin{widetext}
		\begin{equation}\label{E7}
			|S_{\mathrm{tot}}|^{2}
			=
			\left|
			\frac{
				(\omega-\omega_{m1}+i\alpha_0-i\kappa_{m1})
				(\omega-\omega_{m2}+i\beta_0-i\kappa_{m2})
				+\kappa_{m1}\kappa_{m2}
			}{
				(\omega-\omega_{m1}+i\alpha_0+i\kappa_{m1})
				(\omega-\omega_{m2}+i\beta_0+i\kappa_{m2})
				+\kappa_{m1}\kappa_{m2}
			}
			\right|^{2}
			=
			\left|
			\frac{
				(\omega-\tilde{\omega}_{+}')(\omega-\tilde{\omega}_{-}')
			}{
				(\omega-\tilde{\omega}_{+})(\omega-\tilde{\omega}_{-})
			}
			\right|^{2}.
		\end{equation}
	\end{widetext}
	
	Here, $\omega_{m1}$ and $\omega_{m2}$ are the natural resonance frequencies of the two resonators, 
	$\kappa_{m1}$ and $\kappa_{m2}$ are their external coupling rates to the waveguide, and 
	$\alpha_{0}$ and $\beta_{0}$ denote the intrinsic loss rates.
	
	Eq.~(\ref{E7}) also has a zero–pole structure. The poles are
	\begin{equation}\label{E8}
		\begin{aligned}
			\tilde{\omega}_{\pm}
			&= \frac{1}{2}\left[(\omega_{m1}+\omega_{m2})
			- i(\alpha_{0}+\beta_{0}+\kappa_{m1}+\kappa_{m2}) \pm D_{p}\right] \\
			&= \omega_{\pm} - i\gamma_{\pm},
		\end{aligned}
	\end{equation}
	and the zeros of $S_{\mathrm{tot}}$ are
	\begin{equation}\label{E9}
		\begin{aligned}
			\tilde{\omega}_{\pm}'
			&= \frac{1}{2}\left[(\omega_{m1}+\omega_{m2})
			- i(\alpha_{0}+\beta_{0}-\kappa_{m1}-\kappa_{m2}) \pm D_{z}\right] \\
			&= \omega_{\pm}' - i\gamma_{\mathrm{eff},\pm},
		\end{aligned}
	\end{equation}
	where 
	$D_{p}=\sqrt{\left(\Delta - i[(\alpha_{0}+\kappa_{m1})-(\beta_{0}+\kappa_{m2})]\right)^{2}-4\Gamma^{2}}$ and 
	$D_{z}=\sqrt{\left(\Delta - i[(\alpha_{0}-\kappa_{m1})-(\beta_{0}-\kappa_{m2})]\right)^{2}-4\Gamma^{2}}$, 
	with $\Delta=\omega_{m1}-\omega_{m2}$ as the detuning and $\Gamma=\sqrt{\kappa_{m1}\kappa_{m2}}$ as the indirect coupling strength.
	
	The real parts of the poles and zeros correspond to the mode and anti-mode frequencies, while their imaginary parts define the decay rates and the effective decay rates, 
	\begin{equation}\label{E10}
		\gamma_{\pm}
		= \frac{\alpha_{0}+\beta_{0}+\kappa_{m1}+\kappa_{m2}}{2} 
		\mp \frac{1}{2}\operatorname{Im}(D_{p}),
	\end{equation}
	and 
	\begin{equation}\label{E11}
		\gamma_{\mathrm{eff},\pm}
		= \frac{\alpha_{0}+\beta_{0}-\kappa_{m1}-\kappa_{m2}}{2}
		\mp \frac{1}{2}\operatorname{Im}(D_{z}).
	\end{equation}
	
	Under the damping–matching condition \(\alpha_{0}=\kappa_{m1}\) and \(\beta_{0}=\kappa_{m2}\), Eq.~(\ref{E11}) reduces to
	\begin{equation}\label{E12}
		\gamma_{\mathrm{eff},\pm}
		=
		\mp \frac{1}{2}\operatorname{Im}\!\left[
		\sqrt{\Delta^{2}-4\Gamma^{2}}
		\right].
	\end{equation}
	
	For \(|\Delta|/2\Gamma \le 1\), the quantity \(\sqrt{\Delta^{2}-4\Gamma^{2}}\) is purely imaginary and thus \(\gamma_{\mathrm{eff},\pm}\) remains finite. At zero detuning, \(\omega_{m1}=\omega_{m2}=\omega_{0}\), $\Delta=$0, and assuming \(\kappa_{m1}\approx\kappa_{m2}=\kappa\), we obtain \(\gamma_{+}=\kappa\), and \(\gamma_{-}=3\kappa\) and \(\gamma_{\mathrm{eff},\pm}=\mp\kappa\). Eq.~(\ref{E7}) then reduces to
	\begin{equation}\label{E13}
		|S_{\mathrm{tot}}|^{2}
		=
		\frac{
			(\omega-\omega_{0})^{2}+\kappa^{2}
		}{
			(\omega-\omega_{0})^{2}+(3\kappa)^{2}
		},
	\end{equation}
	which exhibits a Lorentzian dip with its FWHM equal to \(2\gamma_{-}=6\kappa\). We note that $\gamma_{\mathrm{eff},\pm}$ can take both positive and negative values; nevertheless, only its magnitude is experimentally relevant since it is directly accessible through the measured inverse spectrum linewidth. Accordingly, we define the FWHM associated with the inverse spectrum as $2|\gamma_{\mathrm{eff},\pm}|$, which yields $2\kappa$ in the present case. At resonance \(\omega=\omega_{0}\), one finds $|S_{\mathrm{tot}}|^{2}=\kappa^{2}/(3\kappa)^{2}\approx 0.1$.
	
	In contrast, when \(|\Delta|/2\Gamma \ge 1\), the expression becomes real and \(\gamma_{\mathrm{eff},\pm}=0\). Consequently,
	
	\begin{equation}\label{E14}
		|S_{\mathrm{tot}}|^{2}
		=
		\left|
		\frac{
			(\omega-\omega_{+}')(\omega-\omega_{-}')
		}{
			(\omega-\tilde{\omega}_{+})(\omega-\tilde{\omega}_{-})
		}
		\right|^{2}.
	\end{equation}
	
	Zeros in the spectrum occur at the two real frequencies $\omega_{\pm}'=\frac{1}{2}[ \omega_{m1}+\omega_{m2} \pm \sqrt{\Delta-4\Gamma^{2}}]$. As detuning increases, both the hybridized modes $\omega_{\pm}$ and the anti-modes $\omega_{\pm}'$ frequencies gradually approach the uncoupled modes; thus, \(|S_{\mathrm{tot}}|^{2}\) can be approximated as the superposition of two Lorentzian dips centered at \(\omega_{m1}\) and \(\omega_{m2}\), with FWHM given by \(2(\alpha_{0}+\kappa_{m1})\) and \(2(\beta_{0}+\kappa_{m2})\), respectively (See Appendix \ref{app:largedetuning}). Similarly, the inverse \(1/|S_{\mathrm{tot}}|^{2}\) can be approximated as the superposition of two Lorentzians with maxima at \(\omega_{m1}\) and \(\omega_{m2}\), whose FWHM vanish.
	
	Fig.~\ref{Fig1}(f) shows a representative spectrum calculated from Eq.~(\ref{E14}) using $\omega_{m1}/2\pi = 6.136~\mathrm{GHz}$, $\omega_{m2}/2\pi = 6.146~\mathrm{GHz}$, $\alpha_{0}/2\pi = \kappa_{m1}/2\pi = 0.8~\mathrm{MHz}$, $\beta_{0}/2\pi = \kappa_{m2}/2\pi = 1.1~\mathrm{MHz}.$ In this spectrum, the decay rates \(\gamma_{\pm}\) determine the FWHM, while the effective decay rates \(\gamma_{\mathrm{eff},\pm}=0\), consistent with the 3~dB-above-minimum linewidth.
	
	\section{Experiment}
	To realize CPA in CMP platforms, the experimental setup is illustrated in Fig.~\ref{Fig1}(d). Ports 1 and 2 are connected to a vector network analyzer (VNA). A signal \( S_T \) is injected into port 1 and is split equally into two symmetric paths—depicted in red and blue—that counter-propagate toward the CMP system as the device under test (DUT), generating inputs with equal power. The outgoing waves propagating to the left and right are extracted by directional couplers and directed to port 2 for independent measurements. The total output power is then obtained by squaring each measured amplitude and adding the corresponding powers.
	
	\begin{figure*}
		\centering
		\includegraphics[width=0.8\linewidth]{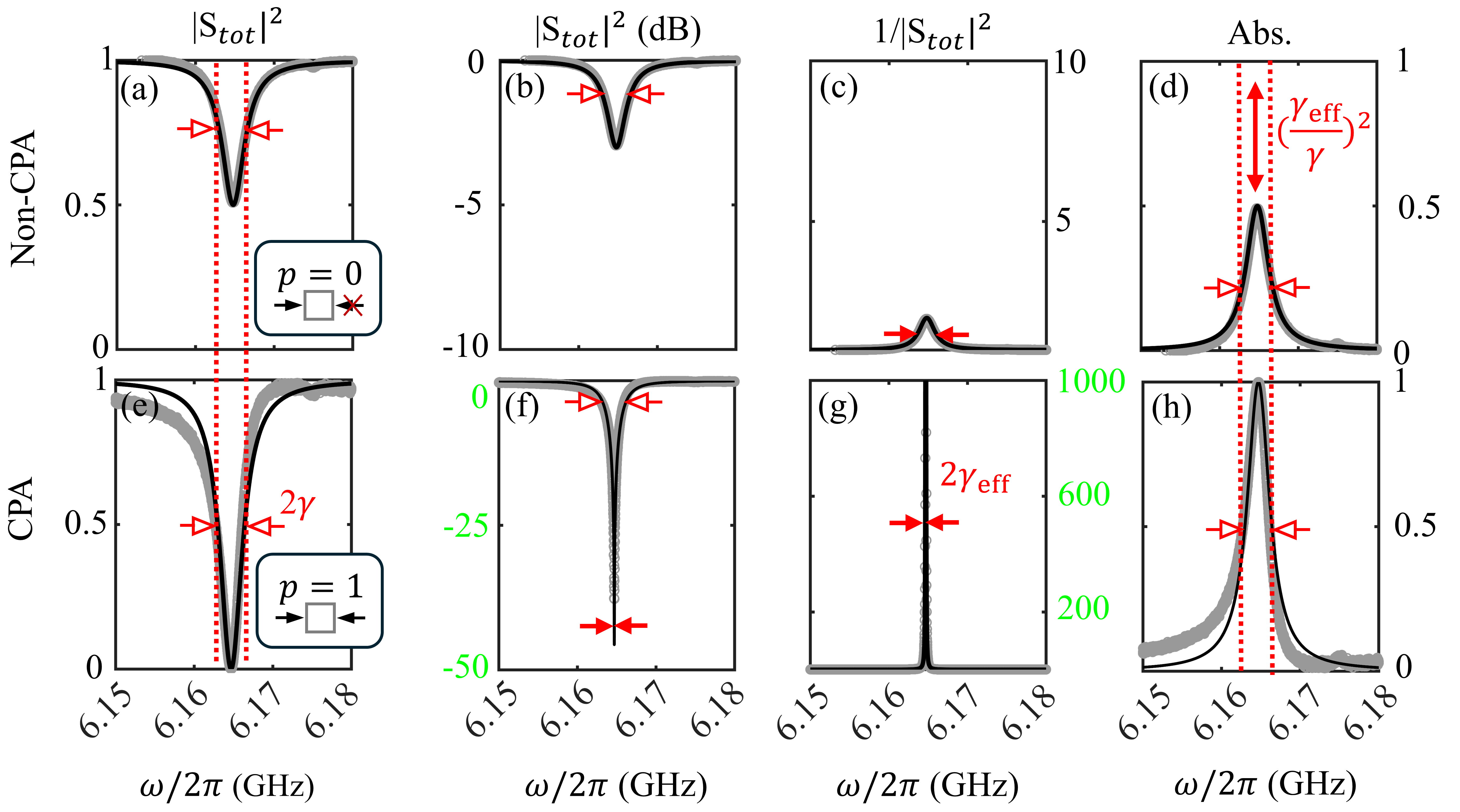}
		\caption{Comparison of spectra for the single resonator system under non-CPA and CPA configurations. Shown are the linear- (a),(e) and dB-scaled (b),(f) total output spectra $|S_{\mathrm{tot}}|^{2}$, the inverse spectra $1/|S_{\mathrm{tot}}|^{2}$ (c),(g), and the absorption spectra (d),(h). Panels (a--d) correspond to the non-CPA case with the $S_{2+}$ port terminated ($p=0$), while panels (e--h) show the CPA case under dual excitation ($p=1$). Experimental data are shown in gray, and theoretical calculations based on Eqs.~(\ref{E6}) and (\ref{E5}) are plotted as black curves. In the non-CPA configuration, $2\gamma$ is extracted from the FWHM of the Lorentzian dip in the linear-scale total-output and absorption spectra (open arrows), which also sets the background response in the dB-scale spectrum.   $2\gamma_{\mathrm{eff}}$ is obtained from the FWHM of the corresponding peak in the inverse spectrum (solid arrows) and remains finite, consistent with the incomplete absorption in (d). Physically, this indicates that part of the input power is not absorbed by the oscillator but remains in the output channel, yielding an output power proportional to $(\gamma_{\mathrm{eff}}/\gamma)^{2}$. Since $\gamma$ is fixed for a given system, tuning $\gamma_{\mathrm{eff}}$ via input control directly regulates the absorption and consequently the output. In the CPA configuration, $\gamma$ remains unchanged, while the total output is fully suppressed and the absorption reaches unity. The zero-output feature is most clearly revealed in the dB-scaled spectrum (f) as an ultra-sharp dip. Correspondingly, the inverse spectrum (g) exhibits a vanishing FWHM, indicating $\gamma_{\mathrm{eff}}=0$, consistent with the zero 3~dB-above-minimum linewidth (solid arrows).}
		\label{fig2a2}
	\end{figure*}
	
	\begin{table}[b]
		\centering
		\begin{ruledtabular}
			\begin{tabular}{lcc}
				\textbf{Systems} & \textbf{Parameters} & \textbf{Value} \\
				\hline
				\multirow{3}{*}{Single YIG} 
				& $\omega_m/2\pi$    & 6.165\,GHz    \\
				& $\alpha/2\pi$    & 0.8\,MHz  \\
				& $\kappa_m/2\pi$    & 0.8\,MHz   \\
				\hline
				\multirow{8}{*}{Coupled YIGs}
				& $\omega_{m1}/2\pi$ & 6.12--6.15\,GHz \\
				& $\omega_{m2}/2\pi$ & 6.136\,GHz \\
				& $\alpha_0/2\pi$    & 0.77\,MHz   \\
				& $\beta_0/2\pi$     & 1.1\,MHz   \\
				& $\kappa_{m1}/2\pi$ & 0.8\,MHz    \\
				& $\kappa_{m2}/2\pi$ & 1.06\,MHz   \\
				& $\Gamma/2\pi$      & 0.89\,MHz   \\
			\end{tabular}
		\end{ruledtabular}
		\caption{\label{tab:params_minimal}
			Key parameters for the single and coupled CMP systems.}
	\end{table}

	\subsection{Single resonator}
	
	To demonstrate single-resonator CPA in a CMP system, a YIG sphere is placed at the center of a coplanar waveguide (CPW), as shown in Fig.~\ref{Fig1}(c). Here, the YIG sphere functions as a magnonic resonator, where the uniform Kittel-mode corresponds to the coherent precession of all spins in the sphere and is therefore widely adopted as a magnetic resonant mode \cite{zhang2014strongly,yang2024anomalous,yaononreciprocity}. First, the YIG sphere is biased by an external magnetic field \( \mathbf{H} \) applied parallel to the direction of signal propagation, which tunes the resonance frequency \( \omega_{m} \) of the Kittel-mode according to
	\begin{equation}\label{htune}
		\frac{\omega_{m}}{2\pi} = \gamma_e \mu_0 \left( |\mathbf{H}| + H_A \right).
	\end{equation}
	
	The vacuum permeability is denoted by \( \mu_0 \), and \( H_A \) represents the effective anisotropy field. From calibration measurements, the gyromagnetic ratio is determined to be \( \gamma_e / 2\pi = 28~\mathrm{GHz/T} \). The intrinsic ($\alpha/2\pi$) and extrinsic ($\kappa_{m}/2\pi$) damping rates of the magnon mode are determined by single-port excitation and fitting the transmission coefficient $S_{21}$ measured at the opposite port. Details of the procedure are provided in Refs.~\cite{lu2025temporal_indirect,yang2024anomalous,yaononreciprocity}.
	
	To fulfill the CPA damping condition \( \alpha = \kappa_{m} \), we adjust the vertical distance between the YIG sphere and the CPW to tune the extrinsic damping, following the procedure described in Ref.~\cite{yaononreciprocity}. A summary of all relevant parameters is provided in Table~\ref{tab:params_minimal}. Additionally, in experimental measurements, the extracted values may depend on the frequency-sweep resolution, defined as the number of sweep points per measurement bandwidth. In this work, a resolution of 100 points per MHz was used.
	
	First, the $S_{2+}$ path is terminated with a 50~$\Omega$ load, such that the system is excited only through $S_{1+}$, rendering the measurement equivalent to a standard transmission-reflection configuration. The measured total output shown in Fig.~\ref{fig2a2}(a) (gray curve) exhibits a single dip with a minimum of approximately 0.5 and a FWHM of 3.2~MHz (open arrows), consistent with $2\gamma = 2(\alpha + \kappa) = 3.2$ MHz. The overall spectrum is well reproduced by the calculation based on Eq.~(\ref{E6}) (black curve) using the parameters listed in Table~\ref{tab:params_minimal}. This result is independently confirmed by the absorption spectrum shown in Fig.~\ref{fig2a2}(d), which yields the same FWHM. In the dB-scaled spectrum [Fig.~\ref{fig2a2}(b)], the apparent FWHM (open arrows) reflects only the broad background response, while analysis of the inverse spectrum $1/|S_{\mathrm{tot}}|^{2}$ shown in Fig.~\ref{fig2a2}(c) yields a FWHM of 2.3~MHz (closed arrows), consistent with $2\gamma_{\mathrm{eff}} = 2\sqrt{\alpha^{2} + \kappa^{2}} = 2\sqrt{2}\,\alpha=2.26$ MHz.
	
	\begin{figure*}
		\centering
		\includegraphics[width=0.8\linewidth]{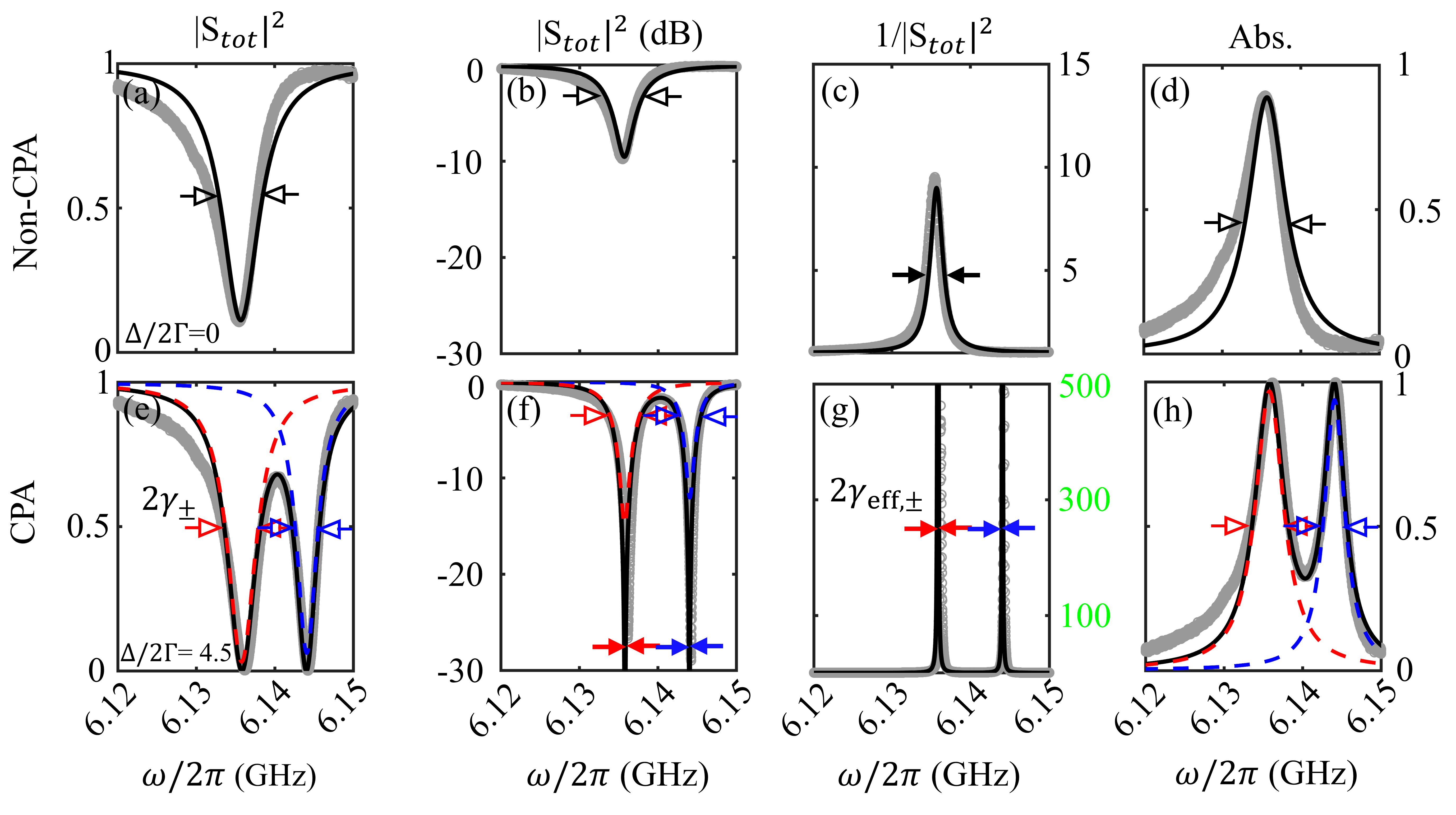}
		\caption{Comparison of spectra for the indirectly coupled system under zero and large detuning. Shown are the linear- (a),(e) and dB-scaled (b),(f) total output spectra $|S_{\mathrm{tot}}|^{2}$, the inverse spectra $1/|S_{\mathrm{tot}}|^{2}$ (c),(g), and the absorption spectra (d),(h). Experimental data are shown in gray, while theoretical results based on Eqs.~(\ref{E13}) and (\ref{E14}) are plotted as black curves. At zero detuning (non-CPA regime), a single Lorentzian dip is observed in the linear-scale total output and absorption spectra, with linewidth $2\gamma_{-}$ extracted from the FWHM (open arrows), which sets the broad background in the dB-scale spectra. At large detuning, the total output exhibits two zero-output dips while the absorption simultaneously reaches unity, indicating CPA. The dips are fitted by two Lorentzians (blue and red dashed lines), yielding linewidths $2\gamma_{+}$ and $2\gamma_{-}$. In contrast, the inverse spectra exhibit vanishing linewidths corresponding to $\gamma_{\mathrm{eff},\pm}=0$, resulting in extremely sharp dips and a zero 3~dB-above-minimum linewidth.}
		\label{fig3}
	\end{figure*}
	
	To demonstrate CPA, the path of $S_{2+}$ is connected such that identical input fields are injected from both directions, with equal amplitude and phase. The measured total output shown in Fig.~\ref{fig2a2}(e) exhibits a dip reaching zero, indicating complete suppression of outgoing signals, while the FWHM remains at $2\gamma = 3.2$~MHz (open arrows), identical to the non-CPA case. The overall spectrum is reproduced by the calculation based on Eq.~(\ref{E5}) using the parameters listed in Table~\ref{tab:params_minimal}. Complete absorption is further confirmed by the absorption spectrum in Fig.~\ref{fig2a2}(h), where the absorption reaches unity. The zero output feature is most clearly highlighted in the total output under the dB-scale shown in Fig.~\ref{fig2a2}(f), where the logarithmic scale magnifies the features associated with small variations near zero output. As a result, an exceptionally sharp dip appears, from which the effective decay rate $\gamma_{\mathrm{eff}}$, estimated using the 3~dB-above-minimum linewidth (closed arrows), vanishes. Analysis of the inverse spectrum further corroborates this behavior, yielding a vanishing FWHM, also corresponding to $2\gamma_{\mathrm{eff}}=0$, in agreement with the discussion following Eq.~(\ref{E5}). 
	
	These observations reveal a fundamental separation between the physical roles of $\gamma$ and $\gamma_{\mathrm{eff}}$ in the single-resonator case. From the frequency-domain perspective, the two parameters manifest themselves through distinct spectral features: the decay rate $\gamma$ is captured by the FWHM extracted from the total-output and absorption spectra, whereas the effective decay rate $\gamma_{\mathrm{eff}}$ is encoded in the FWHM of the inverse spectrum $1/|S_{\mathrm{tot}}|^{2}$. Under the CPA condition, $\gamma_{\mathrm{eff}}=0$, while the intrinsic decay rate $\gamma$ remains unchanged.
	
	Although $\gamma_{\mathrm{eff}}$ formally appears as a linewidth parameter in the inverse spectrum, it does not alter either the total output or the absorption spectra, which are physically meaningful, as both retain a fixed linewidth of $2\gamma$. Instead, decreasing $\gamma_{\mathrm{eff}}$ solely enhances the response magnitude, establishing $\gamma_{\mathrm{eff}}$ as a steady-state amplitude control parameter rather than a true linewidth. A similar interpretation applies to the 3~dB-above-minimum criterion, which simply reflects how close the minimum of $|S_{\mathrm{tot}}|^{2}$ is to zero: a deeper minimum produces a dB-narrower dip without any genuine change in the resonance linewidth.
	
	A complementary time-domain interpretation, detailed in the Appendix~\ref{time1}, further explains this physical distinction. Although both $\gamma$ and $\gamma_{\mathrm{eff}}$ are rates in form, they enter the system dynamics in different ways. The decay rate $\gamma$ appears explicitly in the dynamical equation governing the time evolution of the resonator and, therefore, defines an intrinsic temporal scale of the system, setting the characteristic timescale for both free decay and the driven build-up of the resonator oscillation. This timescale is determined solely by the system parameters and is independent of the input configuration. 
	
	In contrast, the effective decay rate $\gamma_{\mathrm{eff}}$ does not govern the temporal evolution but instead characterizes the degree of destructive interference in the steady state. As a result, $\gamma_{\mathrm{eff}}$ defines an amplitude scale: it controls the steady-state resonator amplitude and the corresponding output level by quantifying the deviation from the maximum achievable resonator amplitude or, equivalently, from the minimum attainable output.
	
	As a consequence, while $\gamma$ remains fixed for a given system, $\gamma_{\mathrm{eff}}$ can be continuously tuned through the input configuration, allowing direct control over the amplitude and output power of the mode at steady-state without altering the intrinsic transient timescale. This tunable amplitude scale is manifested in the frequency domain as the residual, unabsorbed portion of the input power in the absorption spectrum (solid arrows in Fig.~\ref{fig2a2}(d)), and in the time domain as the steady-state output. Together, these results establish $\gamma_{\mathrm{eff}}$ as a controllable amplitude scale that complements the intrinsic temporal scale set by $\gamma$.
	
	\begin{figure}  
		\centering
		\includegraphics[width=\linewidth]{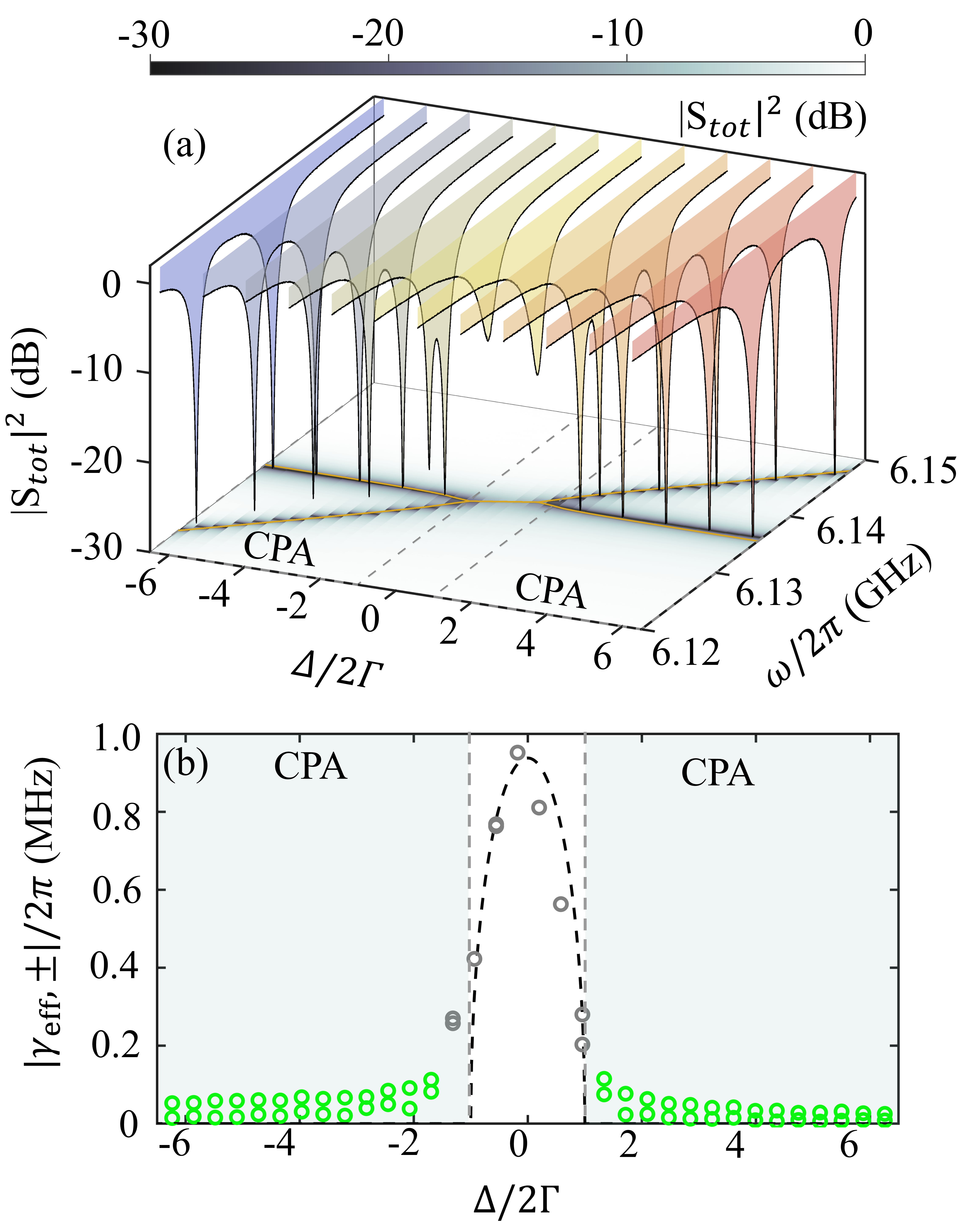}
		\caption{(a) The dB-scale total-output spectra show the evolution of the hybridized system under detuning. Representative spectra and the corresponding color map reveal that the dip positions follow the hybridized anti-mode frequencies (orange curves), indicating level attraction. (b) $\gamma_{\mathrm{eff},\pm}$ extracted from the inverse spectra using Lorentzian fitting. Green and gray circles denote values at $|\Delta/2\Gamma|>1$ and $|\Delta/2\Gamma|<1$, respectively; the dashed curve shows the prediction of Eq.~(\ref{E12}), indicating that $\gamma_{\mathrm{eff},\pm}$ persists over a broad detuning range.}
		\label{fig4}
	\end{figure}

	\subsection{Indirectly coupled resonator}
	To implement CPA in the indirectly coupled CMP system, two $1\,\mathrm{mm}$ YIG spheres are placed on the same CPW, forming two spatially separated resonators. Due to the finite propagation distance between the two coupling points, microwave signals acquire a traveling phase \cite{pozar2012microwave,lu2025temporal_indirect} along the CPW. The traveling phase $\phi$ is tuned to be close to $2\pi$ at the operating frequencies, so the system operates in the dissipative coupling regime, and the counter-propagating inputs reach both YIG spheres with the same phase.

	Both YIG spheres are subjected to the same global bias field $\mathbf{H}$. A small auxiliary coil is placed near one of the spheres. The DC current applied to this coil produces a local tuning field $\Delta H$ that shifts the mode frequency of that YIG sphere, as shown in Fig.~\ref{Fig1}(g). This enables continuous sweeping of the frequency detuning between the two resonators.
	
	The damping-matching procedure follows the same method as used for the single-resonator configuration, ensuring that the damping conditions $\alpha_{0} \approx \kappa_{m1}$ and $\beta_{0} \approx \kappa_{m2}$ are satisfied. In practice, this matching is not exact; therefore, the degree of CPA achieved is limited by the precision of the parameter matching. We note that the extracted parameters may differ slightly from those obtained in the single-YIG calibration because of variations in sample positioning and experimental conditions. A summary of all parameters is provided in Table~\ref{tab:params_minimal}.

	First, by tuning $\Delta H$, we measure the total output spectra in two representative regimes: zero detuning, $|\Delta|/2\Gamma = 0$, and large detuning, $|\Delta|/2\Gamma = 4.5$.

	At zero detuning, the measured total output in Fig.~\ref{fig3}(a) (gray curve) exhibits a single dip with a minimum of approximately 0.1 and a FWHM of 5.5~MHz (open arrows). The overall spectrum is well reproduced by the calculation based on Eq.~(\ref{E13}) (black curve), the observed FWHM is consistent with $2\gamma_{-}/2\pi=5.61$ MHz. This behavior is also confirmed by the absorption spectrum shown in Fig.~\ref{fig3}(d), which yields the same FWHM. By analyzing the inverse spectrum $1/|S_{\mathrm{tot}}|^{2}$ in Fig.~\ref{fig3}(c), we extract a FWHM of 1.8~MHz (closed arrows), close to $2|\gamma_{\mathrm{eff},\pm}|/2\pi=1.79$ MHz.
	
	To demonstrate CPA, the detuning between the magnon modes is increased to $|\Delta|/2\Gamma = 4.5$. The total output measured (gray curve) shown in Fig.~\ref{fig3}(e) exhibits two pronounced dips that reach zero, indicating complete suppression of outgoing signals. Fitting the spectrum with two Lorentzian line shapes (blue and red dashed lines) yields FWHMs of 3.2~MHz and 4.4~MHz, consistent with $2\gamma_{+}=3.13$ MHz and $2\gamma_{-}=4.34$ MHz (open arrows), while the overall spectrum is well reproduced by calculation (black curve) based on Eq.~(\ref{E14}) using the parameters listed in Table~\ref{tab:params_minimal}. The same CPA features are confirmed by the absorption spectrum in Fig.~\ref{fig3}(h), where the absorption peaks reach unity. In the dB-scaled total output shown in Fig.~\ref{fig3}(f), the dips become extremely sharp and the 3~dB-above-minimum linewidth vanishes. Analysis of the inverse spectrum also reveals a vanishing FWHM, corresponding to $2|\gamma_{\mathrm{eff},\pm}| = 0$, in agreement with the discussion following Eq.~(\ref{E14}). A complementary time-domain interpretation of this behavior is provided in Appendix~\ref{double2}. This analysis shows that hybridized modes exhibit distinct decay rates $\gamma_{\pm}$ that set transient timescales, while the CPA condition selectively drives the corresponding $\gamma_{\mathrm{eff},\pm}$ to zero, resulting in complete suppression of the outgoing fields. This time-domain picture is consistent with the spectral features observed in Figs.~\ref{fig3}(f) and (h).
	
	Having established CPA at a fixed detuning and clarified the distinct roles of $\gamma_{\pm}$ and $\gamma_{\mathrm{eff},\pm}$, we now examine the spectral evolution of the system under detuning. Fig.~\ref{fig4}(a) presents representative dB-scale spectra of $|S_{\mathrm{tot}}|^{2}$ together with a two-dimensional color map. The spectral dip positions closely follow the hybridized anti-mode frequencies predicted by the real part of Eq.~(\ref{E9}) (orange curves), thereby demonstrating the characteristic level-attraction dispersion of the anti-modes. Notably, this result indicates that the absorption maxima also occur at the anti-mode frequencies. This behavior arises from a parameter-matching condition different from that considered in previous studies, where absorption maxima were shown to indicate mode frequencies \cite{wang_2024, lu2025temporal_indirect}. More details are provided in Appendix~\ref{abs}.

	To quantitatively examine the perfect-absorption condition under detuning, we extract the effective damping rates $\gamma_{\mathrm{eff},\pm}$ from the inverse spectra $1/|S_{\mathrm{tot}}|^{2}$ using a Lorentzian fitting procedure. The extracted values are summarized in Fig.~\ref{fig4}(b), where the data points obtained for $|\Delta/2\Gamma|>1$ are shown as green circles and those for $|\Delta/2\Gamma|<1$ are shown as gray circles. These results are consistent with the dashed curve, which represents the theoretical prediction given by Eq.~(\ref{E12}).
	
	These results reveal two key distinctions between indirect and direct coupling under CPA~\cite{shen2025polaromechanics,zhang2017exceptional}. Indirect coupling leads to level attraction of the anti-mode frequencies, whereas direct coupling exhibits level repulsion under detuning. Moreover, CPA persists over a broad detuning range for indirect coupling but occurs only at a single detuning point for direct coupling.
	
	The magnetically controlled detuning window that supports CPA (shaded area in Fig.~\ref{fig4}(b)) highlights a key advantage of indirect coupling: the absorption frequency can be tuned directly by the magnetic field, without requiring control over the spatial overlap of the mode profiles inherent to direct coupling. This magnetic tunability enables reconfigurable dual-frequency microwave absorbers.
	
	\begin{figure}
		\centering
		\includegraphics[width=\linewidth]{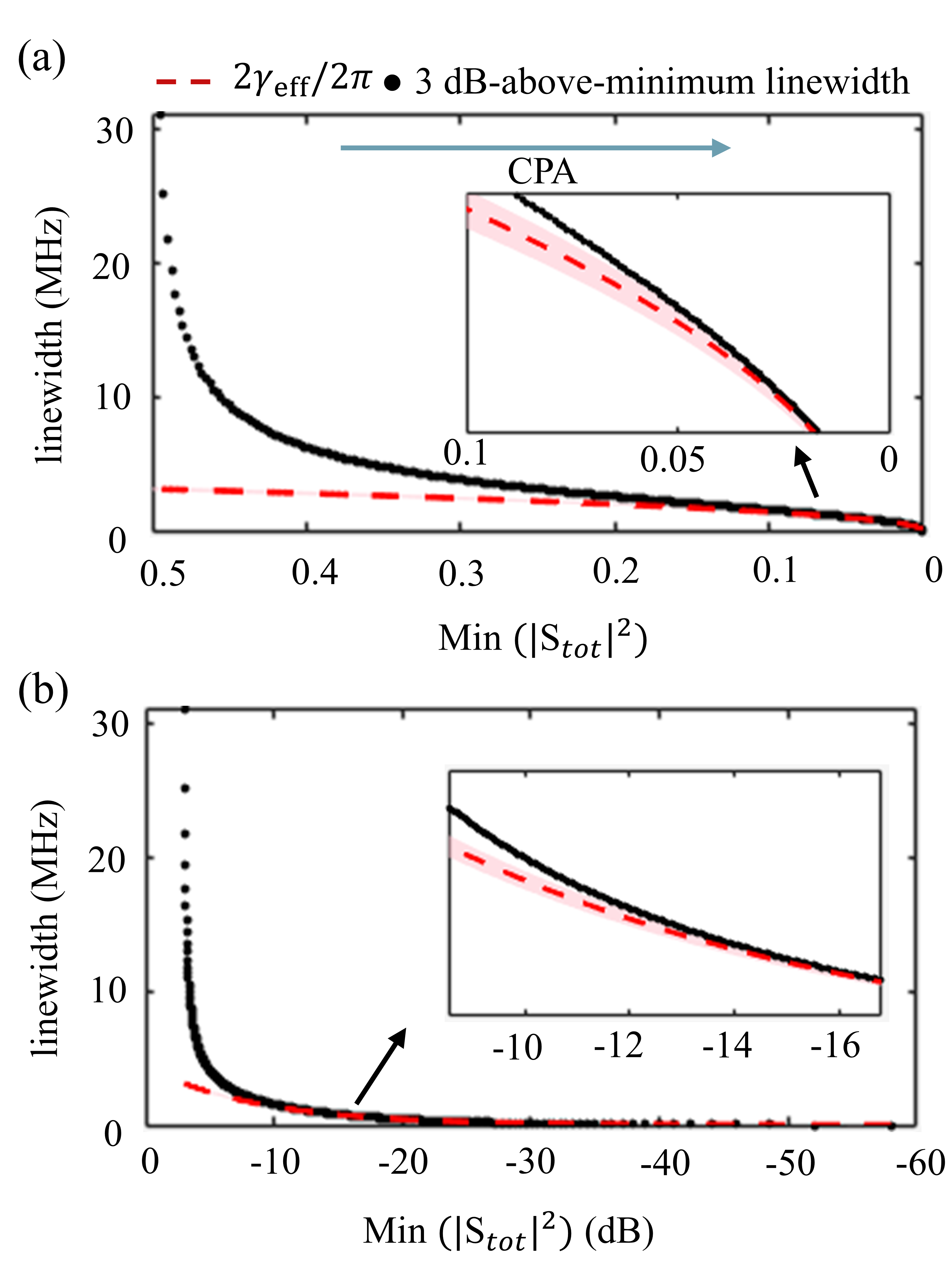}
		\caption{ Single-resonator case under matched damping $(\alpha=\kappa)$: By sweeping $p$, we tune $|S_{tot}(\omega=\omega_{m})|^2$ from Eq.~(\ref{F2}) and examine the calculated $2\gamma_{\mathrm{eff}}$ (red dashed line) as well as the 3~dB-above-minimum linewidth obtained from Eq.~(\ref{F3}) (black dots). The pink shaded region indicates a $\pm5\%$ deviation of $2\gamma_{\mathrm{eff}}$. The results are presented in (a) on a linear scale and in (b) on the dB-scale. It is clear that $2\gamma_{\mathrm{eff}}$ and the 3~dB-above-minimum linewidth overlap only when the total output under 0.05; beyond this point, the two definitions deviate significantly.}
		\label{fig8}
	\end{figure}
	
	\section{conclusion}
	In conclusion, we have experimentally and theoretically established a clear distinction between the decay rate $\gamma$ and the effective decay rate $\gamma_{\mathrm{eff}}$ in CMP systems under CPA. Through pole--zero analysis and spectral measurements, we demonstrate that $\gamma$ is reflected in the FWHM of the total-output and absorption spectra. In contrast, the effective decay rate $\gamma_{\mathrm{eff}}$ reflects the FWHM of the inverse spectrum and is visually associated with the narrow spectral dip observed under CPA. Importantly, $\gamma_{\mathrm{eff}}$ does not represent a physical decay rate or linewidth; instead, it governs the total output amplitude. While $\gamma$ remains finite, $\gamma_{\mathrm{eff}}$ can approach zero under CPA. Furthermore, CPA under indirect coupling exhibits two spectral features absent in directly coupled systems: anti-mode level attraction induced by frequency detuning and broadband perfect absorption, rather than being restricted to a single operating point. These findings not only complete the picture of CPA in CMP platforms but also open avenues for magnetically reconfigurable absorbers and other tunable non-Hermitian magnonic devices.
	
	\paragraph{Note added.}—While this manuscript was in preparation, an interesting and related work by Ebrahimi et al. appeared on arXiv, examining CPA spectral features in both dB and linear representations and discussing the associated linewidth in the context of direct coupling \cite{Jdavis}.
	
	\section{Acknowledgments}
	This work has been funded by NSERC Discovery Grants and NSERC Discovery Accelerator Supplements (C.-M. H.). We thank Jian-Qiang You, Jie Li, Zi-Qi Wang, and Ryan McKenzie for discussions. 
	
	\section{DATA AVAILABILITY}
	The data that support the findings of this article are openly
	available \cite{dataavilibility}.

	\medskip
	\appendix

	\section{Derivation of $|S_{tot}|^2$ for a single resonator CPA} \label{single}
	In this Appendix, we derive the normalized total-output spectrum $|S_{\mathrm{tot}}|^{2}$ for a standing-wave resonator side-coupled to a single waveguide supporting two input channels. Here, the waveguide supports traveling waves propagating in opposite directions, so the input and output fields are naturally defined at two spatially separated ports.
	
	In the present traveling-wave configuration, the resonator is simultaneously driven by the inputs from both ports. Using TCMT~\cite{yang2024anomalous}, the resonator amplitude $\hat{m}(t)$ obeys
	\begin{equation}
		\frac{d\hat{m}}{dt}
		= -i\big(\omega_{m} - i\alpha - i\kappa_{m}\big)\hat{m}
		- i\sqrt{\kappa_{m}}\,S_{1+}
		- i\sqrt{\kappa_{m}}\,S_{2+},
		\label{A1}
	\end{equation}
	where $\omega_{m}$ is the resonance frequency, $\alpha$ is the intrinsic decay rate, and $\kappa_{m}$ is the external coupling rate. The inputs at ports 1 and 2 are denoted by $S_{1+}$ and $S_{2+}$.
	
	The input–output relations are \cite{yang2024anomalous}
	\begin{align}
		S_{1-} &= S_{2+}-i\sqrt{\kappa_{m}}\hat{m}, \label{A2}\\
		S_{2-} &= S_{1+}-i\sqrt{\kappa_{m}}\hat{m}. \label{A3}
	\end{align}
	
	Under a coherent drive at frequency $\omega$, we adopt the steady-state ansatz $\hat{m}(t) = m_{0} e^{-i\omega t}$. Thus, in steady-state, 
	\begin{equation}
		\hat{m}
		=\frac{\sqrt{\kappa_{m}}\bigl(S_{1+}+S_{2+}\bigr)}
		{\omega-\omega_{m}+i(\alpha+\kappa_{m})}.
		\label{A4}
	\end{equation}
	
	We impose the amplitude relation $|S_{2+}| = p|S_{1+}|$, where $p$ denotes the input-amplitude ratio. Substituting Eqs.(\ref{A4}) into Eqs.(\ref{A2})–(\ref{A3}), steady-state output fields can be expressed solely in terms of $S_{1+}$. The total output power $|S_{1-}|^{2} + |S_{2-}|^{2}$, normalized by the total input power $|S_{1+}|^{2} + |S_{2+}|^{2}$, then yields
	
	\begin{widetext}
		\begin{equation}
			|S_{\mathrm{tot}}|^{2}
			=\frac{|S_{1-}|^{2}+|S_{2-}|^{2}}
			{|S_{1+}|^{2}+|S_{2+}|^{2}}
			=\frac{|S_{1-}|^{2}+|S_{2-}|^{2}}
			{(1+p^{2})|S_{1+}|^{2}}=
			\frac{(\omega-\omega_{m})^{2} + \alpha^{2} + \kappa_{m}^{2}
				-\dfrac{4p\alpha\kappa_{m}}{1+p^{2}}}
			{(\omega-\omega_{m})^{2}+(\alpha+\kappa_{m})^{2}}.
			\label{A8}
		\end{equation}
	\end{widetext}
	
	\section{Comparison between $\gamma_{\mathrm{eff},\pm}$ and the 3~dB-above-minimum linewidth} \label{comparison}
	
	To visually estimate $\gamma_{\mathrm{eff}}$ from $|S_{\mathrm{tot}}|^{2}$ on the dB-scale, we define the width at 3~dB above the minimum as the 3~dB-above-minimum linewidth. This approach is appropriate under the CPA condition, where the effective decay rate vanishes and the resonance dip becomes infinitely sharp, leading to a zero 3~dB-above-minimum linewidth. However, for non-CPA cases with a finite effective decay rate, the range of validity of this phenomenological definition needs to be carefully examined.
	
	In this Appendix, we compare $\gamma_{\mathrm{eff}}$ with the linewidth extracted from the 3~dB-above-minimum criterion for both single-resonator and indirectly coupled resonator systems, and identify the parameter regime in which this phenomenological linewidth provides a reliable approximation.
	
	For a single resonator under matched damping conditions $\alpha=\kappa$,
	Eq.~(\ref{E1}) reduces to
	\begin{equation}\label{F1}
		|S_{\mathrm{tot}}|^{2}=
		\frac{
			(\omega-\omega_{m})^{2}
			+2\kappa_{m}^{2}
			-\dfrac{4 p \kappa_{m}^2}{1+p^{2}}
		}{
			(\omega-\omega_{m})^{2}
			+(2\kappa_{m})^{2}
		}.
	\end{equation}

	The minimum of $|S_{\mathrm{tot}}|^{2}$ occurs at $\omega=\omega_{m}$, with a value of
	\begin{equation}
		|S_{\mathrm{tot}}(\omega=\omega_{m})|^{2}
		=
		\frac{
			2\kappa_{m}^{2}
			-\dfrac{4 p \kappa_{m}^2}{1+p^{2}}
		}{
			(2\kappa_{m})^{2}
		}.
	\end{equation}
	
	At 3 dB above the minimum, the corresponding spectral value is given by
	\begin{equation}\label{F2}
		|S_{\mathrm{tot}}|_{3\mathrm{dB}}^{2}
		=
		2|S_{\mathrm{tot}}(\omega=\omega_{m})|^{2}
		=
		\frac{
			4\kappa_{m}^{2}
			-\dfrac{8 p\kappa_{m}^2}{1+p^{2}}
		}{
			(2\kappa_{m})^{2}
		}.
	\end{equation}
	
	Substituting Eq.~(\ref{F2}) into Eq.~(\ref{F1}) and solving for $\omega$, the
	resulting 3~dB-above-minimum linewidth is obtained as
	\begin{equation}\label{F3}
		\Delta\omega_{3\mathrm{dB}}
		=
		2\kappa_m |p-1|
		\sqrt{\frac{1}{p}}.
	\end{equation}
	
	This expression shows that $\Delta\omega_{3\mathrm{dB}}$ diverges as
	$p\rightarrow 0$, while it vanishes in the CPA limit $p\rightarrow 1$,
	consistent with the emergence of an infinitely sharp absorption dip. By sweeping
	$p$ from 0 to 1, Fig.~\ref{fig8} compares the 3~dB-above-minimum linewidth given by
	Eq.~(\ref{F3}) (black dots) with the effective decay rate
	$2\gamma_{\mathrm{eff}}$ calculated from Eq.~(\ref{E3}) (red dashed line),
	using the parameters listed in Table~\ref{tab:params_minimal} under different minimum outputs in both linear (a) and the dB-scale (b). Away from the CPA condition ($|S_{tot}(\omega=\omega_{m})|^2>$0.05 (above -12 dB)), noticeable deviations
	appear, and the 3~dB-above-minimum linewidth exceeds $2\gamma_{\mathrm{eff}}$ by more than
	$5\%$. In contrast, closer to the CPA regime ($|S_{tot}(\omega=\omega_{m})|^2<$0.05 (below -12 dB)), the discrepancy between the two definitions of linewidth remains below $5\%$, indicating that the 3~dB-above-minimum linewidth provides a reasonable approximation to
	$2\gamma_{\mathrm{eff}}$.
	
	\section{Derivation of $|S_{tot}|^2$ for indirectly coupled resonators}\label{double}
	Here we derive the expression for the normalized total-output spectrum $|S_{\mathrm{tot}}|^{2}$ for two indirectly coupled standing-wave resonators side-coupled to a waveguide. Following the matrix formulation developed in Ref.~\cite{yang2024anomalous}, we focus on the purely dissipative coupling regime and therefore assume a traveling phase $\phi=2\pi$, together with identical coupling of each resonator to the forward and backward-propagating waves in the waveguide. Under these conditions, the temporal coupled-mode equations take the form of
	\begin{widetext}
		\begin{equation}
			i\frac{d}{dt}
			\begin{pmatrix}
				\hat{m}_2 \\[2pt] \hat{m}_1
			\end{pmatrix}
			=
			\begin{pmatrix}
				\omega_{m2}-i(\beta_0+\kappa_{m2}) & -i\Gamma \\
				-i\Gamma & \omega_{m1}-i(\alpha_0+\kappa_{m1})
			\end{pmatrix}
			\begin{pmatrix}
				\hat{m}_2 \\[2pt] \hat{m}_1
			\end{pmatrix}
			+
			\begin{pmatrix}
				\sqrt{\kappa_{m2}} & \sqrt{\kappa_{m2}} \\
				\sqrt{\kappa_{m1}} & \sqrt{\kappa_{m1}}
			\end{pmatrix}
			\begin{pmatrix}
				S_{1+} \\[2pt] S_{2+}
			\end{pmatrix},
			\label{B1}
		\end{equation}
	\end{widetext}
	where $\hat{m}_{1,2}$ denote the amplitudes of the resonator mode, $\omega_{m1,m2}$ are the resonance frequencies, $\alpha_0$ and $\beta_0$ are the intrinsic loss rates, $\kappa_{m1,m2}$ the external coupling rates, and $\Gamma=\sqrt{\kappa_{m1}\kappa_{m2}}$ is the indirect coupling strength.
	
	The input-output relations are~\cite{yang2024anomalous}
	\begin{align}
		S_{2-} &= S_{1+} - i\sqrt{\kappa_{m1}}\,\hat{m}_1 - i\sqrt{\kappa_{m2}}\,\hat{m}_2, \label{B2}\\
		S_{1-} &= S_{2+} - i\sqrt{\kappa_{m2}}\,\hat{m}_2 - i\sqrt{\kappa_{m1}}\,\hat{m}_1. \label{B3}
	\end{align}
	
	In steady-state, we adopt the ansatz $\hat{m}_{1,2}(t) = m_{1,2} e^{-i\omega t}$. Solving the coupled-mode equations then yields the steady-state amplitudes
	\begin{widetext}
		\begin{equation}
			\begin{pmatrix}
				m_2 \\[2pt] m_1
			\end{pmatrix}
			=
			\begin{pmatrix}
				\omega-\omega_{m2}+i(\beta_0+\kappa_{m2}) & i\Gamma \\
				i\Gamma & \omega-\omega_{m1}+i(\alpha_0+\kappa_{m1})
			\end{pmatrix}^{-1}
			\begin{pmatrix}
				\sqrt{\kappa_{m2}} & \sqrt{\kappa_{m2}} \\
				\sqrt{\kappa_{m1}} & \sqrt{\kappa_{m1}}
			\end{pmatrix}
			\begin{pmatrix}
				S_{1+} \\[2pt] S_{2+}
			\end{pmatrix}.
			\label{B4}
		\end{equation}
	\end{widetext}
	
	Assuming coherent excitation with equal amplitudes, $S_{2+} = S_{1+}$, and substituting Eq.~(\ref{B4}) into Eqs.(\ref{B2}) and (\ref{B3}), the output fields can be expressed as functions of $S_{1+}$. Evaluating the total output power $|S_{1-}|^{2}+|S_{2-}|^{2}$ and normalizing it by the total input power $|S_{1+}|^{2}+|S_{2+}|^{2}$ yields
	\begin{widetext}
		\begin{equation}
			|S_{\mathrm{tot}}|^{2}
			=\frac{|S_{1-}|^{2}+|S_{2-}|^{2}}{2|S_{1+}|^{2}}
			=
			\left|
			\frac{
				(\omega-\omega_{m1}+i\alpha_0-i\kappa_{m1})
				(\omega-\omega_{m2}+i\beta_0-i\kappa_{m2})
				+\kappa_{m1}\kappa_{m2}
			}{
				(\omega-\omega_{m1}+i\alpha_0+i\kappa_{m1})
				(\omega-\omega_{m2}+i\beta_0+i\kappa_{m2})
				+\kappa_{m1}\kappa_{m2}
			}
			\right|^{2}.
			\label{B5}
		\end{equation}
	\end{widetext}

	\section{Total output under large-detuning limit: double lorentz lineshape}
	\label{app:largedetuning}
	
\begin{figure}
		\centering
		\includegraphics[width=\linewidth]{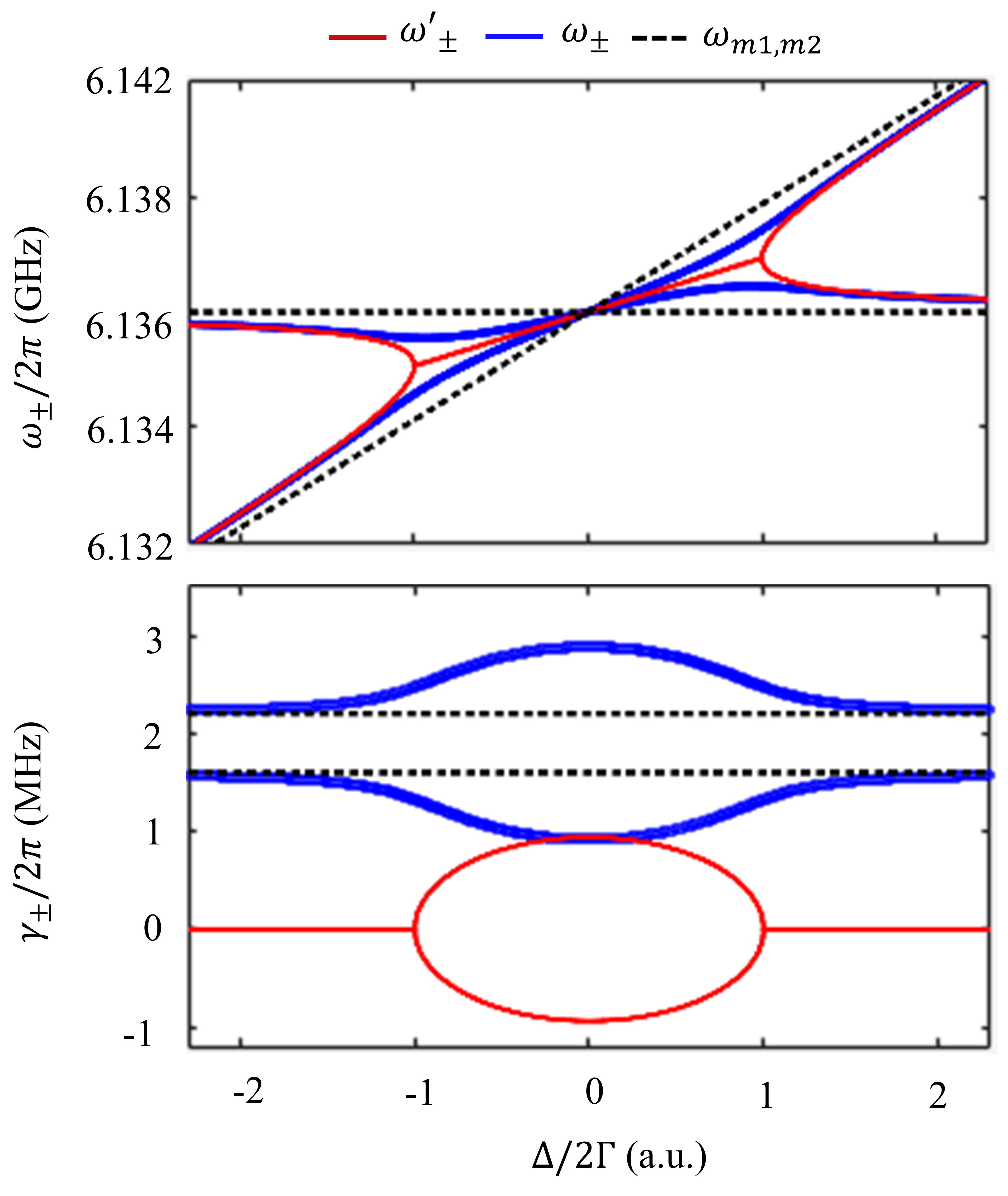}
		\caption{Calculated the hybridized modes (blue) and anti-mode (red) and uncoupled resonators (dashed lines) in (a) their frequencies, (b) damping rates.}
		\label{fig7}
	\end{figure}

	In this Appendix, we provide the theoretical background for the spectral response of the indirectly coupled system in the large-detuning regime. In this regime, where the frequency separation between the two resonators is much larger than their respective damping rates, the hybridization between the two modes becomes weak. Fig.~\ref{fig7} illustrates this behavior using the parameters in the Table. \ref{tab:params_minimal}, while sweeping the detuning $\Delta$ between the two resonators. As the detuning increases, the hybridized modes (blue) and anti-mode frequencies (red) gradually approach those of the uncoupled resonator frequencies (dashed lines), as shown in Fig.~\ref{fig7}(a).
	
	In this limit,
	$\omega_{+}\approx \omega_{+}' \approx \omega_{m1}$, and $\omega_{-} \approx \omega_{-}' \approx \omega_{m2}$. At the same time, the corresponding decay rates $\gamma_{\pm}$ approach those of the uncoupled resonators, while the effective decay rates $\gamma_{\mathrm{eff},\pm}$ remain zero, as shown in Figs.~\ref{fig7}(a) and (b).
	
	Under these conditions, Eq.~(\ref{E14}) reduces to
	\begin{small}
		\begin{widetext}
			\begin{align}
				|S_{\mathrm{tot}}|^{2}
				&=
				\left(
				1-\frac{(\alpha_{0}+\kappa_{m1})^{2}}
				{(\omega-\omega_{m1})^{2}+(\alpha_{0}+\kappa_{m1})^{2}}
				\right)
				\left(
				1-\frac{(\beta_{0}+\kappa_{m2})^{2}}
				{(\omega-\omega_{m2})^{2}+(\beta_{0}+\kappa_{m2})^{2}}
				\right) \notag \\
				&=
				1
				-\frac{(\alpha_{0}+\kappa_{m1})^{2}}
				{(\omega-\omega_{m1})^{2}+(\alpha_{0}+\kappa_{m1})^{2}}
				-\frac{(\beta_{0}+\kappa_{m2})^{2}}
				{(\omega-\omega_{m2})^{2}+(\beta_{0}+\kappa_{m2})^{2}}
				+\frac{(\alpha_{0}+\kappa_{m1})^{2}}
				{(\omega-\omega_{m1})^{2}+(\alpha_{0}+\kappa_{m1})^{2}}
				\frac{(\beta_{0}+\kappa_{m2})^{2}}
				{(\omega-\omega_{m2})^{2}+(\beta_{0}+\kappa_{m2})^{2}} .
			\end{align}
		\end{widetext}
	\end{small}
	
	In the large-detuning limit, the damping rates are much smaller than the frequency separation between the two resonances. Consequently, the cross term in the last line becomes negligible, and the total-output spectrum reduces to
	\begin{widetext}
		\begin{equation}
			\label{EE2}
			|S_{\mathrm{tot}}|^{2}
			\approx
			1
			-\frac{(\alpha_{0}+\kappa_{m1})^{2}}
			{(\omega-\omega_{m1})^{2}+(\alpha_{0}+\kappa_{m1})^{2}}
			-\frac{(\beta_{0}+\kappa_{m2})^{2}}
			{(\omega-\omega_{m2})^{2}+(\beta_{0}+\kappa_{m2})^{2}} .
		\end{equation}
	\end{widetext}
	
	Eq.~(\ref{EE2}) shows that, in this regime, the total-output spectrum is well described by the superposition of two independent Lorentzian dips centered at $\omega_{m1}$ and $\omega_{m2}$. The corresponding FWHMs are given by $2(\alpha_{0}+\kappa_{m1})$ and $2(\beta_{0}+\kappa_{m2})$, respectively.
	
	\section{Time-domain interpretation of $\gamma$ and $\gamma_{\mathrm{eff}}$ for a single resonator}\label{time1}
	\begin{figure*}
		\centering
		\includegraphics[width=0.9\linewidth]{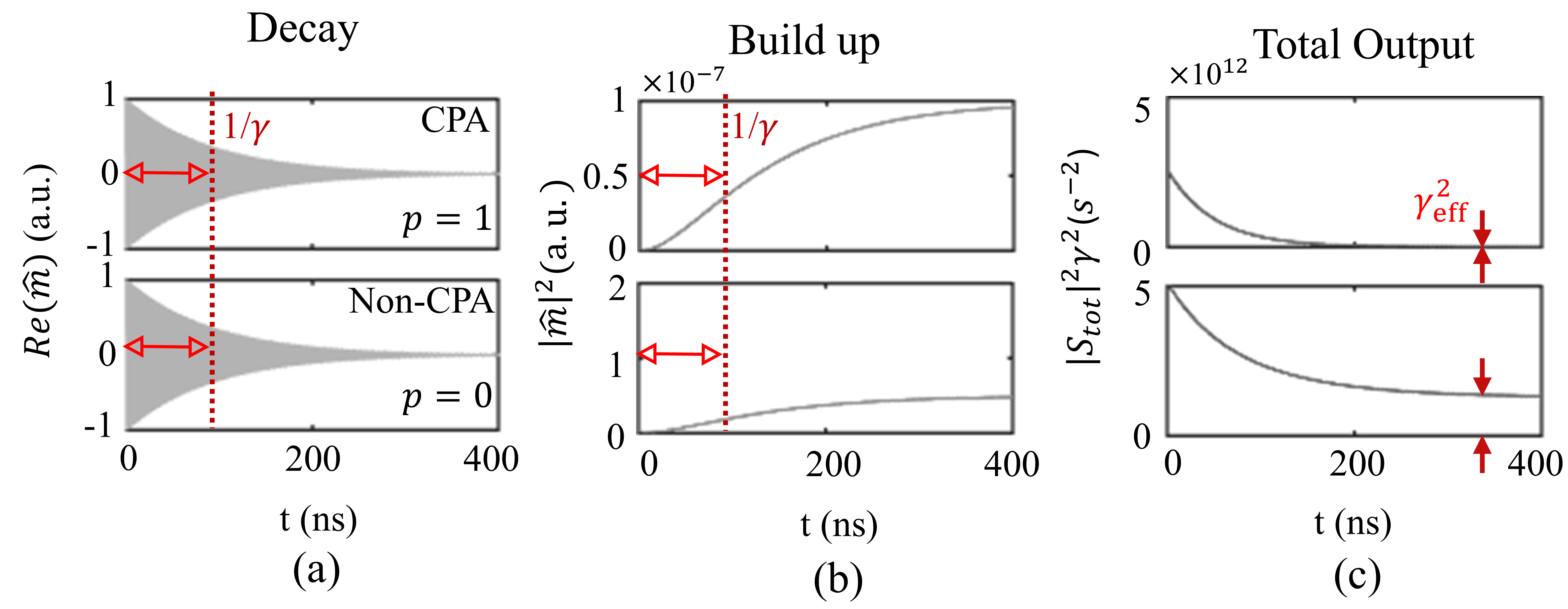}
		\caption{Calculated time-domain dynamics: (a) decay, (b) build up, and (c) normalized total output power. The decay rate $\gamma$ defines the decay and build up time (red open arrows), while $\gamma_{\mathrm{eff}}$ is reflected in the steady-state output and resonator power (blue solid arrows).}
		\label{fig3a2}
	\end{figure*}
	In this appendix, we provide a time-domain analysis that complements the
	frequency-domain pole--zero picture discussed in the main text. Although
	$\gamma$ and $\gamma_{\mathrm{eff}}$ are defined through the imaginary parts
	of the pole and zero, respectively, their dynamical roles become transparent
	in the time domain: $\gamma$ governs the transient decay and build-up of the
	resonator field, whereas $\gamma_{\mathrm{eff}}$ determines the steady-state
	output level and vanishes under the CPA condition.
	
	\paragraph{Free decay}
	To characterize the free evolution of the system, we set $S_{1+}=S_{2+}=0$ in Eq.~(\ref{A1}). Assuming the initial condition $\hat m(t=0)=m_0$, we obtain the transient decay solution
	\begin{equation}\label{C1}
		\hat m(t)
		=
		m_0\,e^{-i\omega_{m} t-(\alpha+\kappa_{m})t},
	\end{equation}
	showing that the field decays exponentially at the rate
	$\gamma=\alpha+\kappa_{m}$, with a decay time $\tau=1/\gamma$, as demonstrated
	in Fig.~\ref{fig3a2}(a) (open red arrows) with parameters: $\omega_{m}/2\pi = $6.165~GHz, $\alpha/2\pi = $0.856~MHz, $\kappa_{m}/2\pi = $0.865~MHz.
	
	\paragraph{Build-up}
	Under continuous driving from both ports, $S_{1+}=|S_{1+}|e^{-i\omega t}$ and $S_{2+}=|S_{2+}|e^{-i\omega t}$ with $|S_{2+}|=p|S_{1+}|$, the transient solution of Eq.~(\ref{A1}) starting from rest is
	\begin{equation}\label{C2}
		\hat m(t)
		=
		\frac{\sqrt{\kappa_{m}} S_{1+}(1+p)}
		{\omega-\omega_{m}+i(\alpha+\kappa_{m})}
		\left[1-e^{i(\omega-\omega_{m}) t}e^{-(\alpha+\kappa_{m}) t}\right].
	\end{equation}
	
	At resonance, $\omega=\omega_{m}$, the amplitude builds up as
	$(1-e^{-\gamma t})$, indicating the characteristic timescale $1/\gamma$, independent of $p$, as shown in Fig.~\ref{fig3a2}(b) (open red arrows). Thus, the parameter $\gamma$ sets the
	universal transient timescale for both decay and driven build-up.
	
	\paragraph{Steady state}
	In steady-state, the total output $|S_{tot}|^2=(|S_{1-}|^2+|S_{2-}|^2)/(|S_{1+}|^2+|S_{2+}|^2)=(|S_{1-}|^2+|S_{2-}|^2)/((1+p)^2|S_{1+}|^2)$, can be obtained by combining Eqs.(\ref{A2}), (\ref{A3}), and (\ref{C2}) in the limit $t\rightarrow\infty$. The resulting total output from the system under $\omega=\omega_{m}$, is
	\begin{equation}
		|S_{tot}|^{2}=\gamma_{\mathrm{eff}}^{2}/\gamma^{2}.
	\end{equation}
	
	For a given system, both $\alpha_{0}$ and $\kappa_{m}$ are fixed, and therefore $\gamma$ is also fixed. This expression thus shows that $\gamma_{\mathrm{eff}}$ directly determines the steady-state output level: it vanishes under the CPA condition ($\gamma_{\mathrm{eff}}=0$) and becomes finite under non-CPA conditions, as illustrated by the solid arrows in Fig.~\ref{fig3a2}(c).
	
	In summary, the time-domain analysis provides a clear dynamical interpretation of the two rates. The intrinsic decay rate $\gamma$ sets a universal temporal scale for the system, governing the timescale of both free decay and build up of the resonator oscillation, independent of the input configuration. In contrast, the effective decay rate $\gamma_{\mathrm{eff}}$ does not affect transient dynamics but instead controls the steady-state response by determining the output level. Under the CPA condition, $\gamma_{\mathrm{eff}}$ vanishes, leading to maximal steady-state energy storage in the resonator and zero output. This separation highlights the complementary roles of $\gamma$ as a temporal scale and $\gamma_{\mathrm{eff}}$ as an amplitude scale in system dynamics.
	
	\section{Time-domain interpretation of $\gamma_{\pm}$ and 
		$\gamma_{\mathrm{eff},\pm}$ for indirectly coupled resonators}\label{double2}
	
	In this appendix, we extend the time-domain analysis to the indirectly coupled two-resonator system. Due to hybridization, dynamics are governed by two eigenmodes with decay rates $\gamma_{\pm}$ and two associated anti-modes characterized by effective decay rates $\gamma_{\mathrm{eff},\pm}$.
	\paragraph{Free decay}
	To characterize the free evolution of the system, we set $S_{1+}=S_{2+}=0$ in Eq.~(\ref{B1}), which yields the transient solution
	\begin{equation}\label{D1}
		\hat m_{1,2}(t)=
		A_{+}^{(m_{1,2})} e^{-\gamma_{+} t} e^{-i\omega_{+} t}
		+
		A_{-}^{(m_{1,2})} e^{-\gamma_{-} t} e^{-i\omega_{-} t},
	\end{equation}
	where $A_{\pm}^{(m_{1,2})}$ are the amplitude coefficients determined by the initial conditions. This expression shows that the dynamics of each physical resonator are given by a superposition of two hybrid eigenmodes with mode frequencies $\omega_{\pm}$ and damping rates $\gamma_{\pm}$.
	
\begin{figure*}
		\centering
		\includegraphics[width=\linewidth]{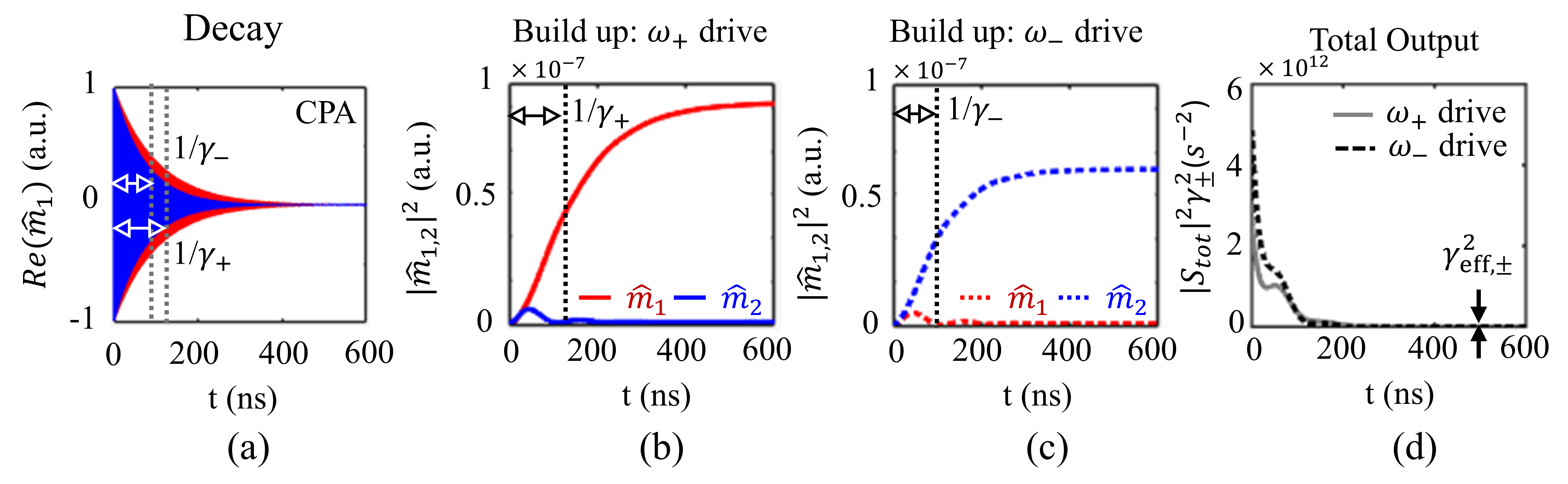}
		\caption{Calculated time-domain dynamics: (a) decay, (b) build-up, and
			(c) normalized total output power. The decay rate $\gamma_{\pm}$ defines
			the decay and build-up times (open arrows), while $\gamma_{\mathrm{eff,\pm}}$
			is reflected in the steady-state outputs (solid arrows).}
		\label{fig3a3}
	\end{figure*}

	Setting $A_{+(-)}^{(m_{1,2})}=0$, the temporal envelope decays at a rate $\gamma_{-(+)}$, resulting in decay times $\tau_{\pm}=1/\gamma_{-(+)}$. This behavior is illustrated in Fig.~\ref{fig3a3}(a), where the open arrows indicate the decay times $\tau_{\pm}$. The calculations are performed using the parameters $\omega_{m1}/2\pi = 6.146$~GHz, $\omega_{m2}/2\pi = 6.136$~GHz, $\alpha_{0}/2\pi = \kappa_{m1}/2\pi = 0.8$~MHz, and $\beta_{0}/2\pi = \kappa_{m2}/2\pi = 1$~MHz.
	
	\paragraph{Build up}
	In hybridized systems, two eigenmode frequencies can coexist in the dynamics, and probing at different frequencies leads to different modal weightings in the response. Here, we consider the simple case of identical excitation from both ports at the eigenfrequency $\omega_{+}$, such that only the $\omega_{+}$ mode is selectively excited. Under large detuning, the difference between $\omega_{+}$ and $\omega_{-}$  becomes significant; therefore, the excitation of the $\omega_{-}$ mode can be neglected. By solving Eq.~(\ref{B1}), the build up transient is 
	\begin{widetext}
		\begin{equation}\label{D2}
			\hat m_{1(2)}(t)
			=
			\frac{
				\sqrt{\kappa_{m1(2)}}
				\bigl(\omega-\omega_{m2(1)} + i\beta_{0}(\alpha_{0})\bigr)
				(S_{1+}+S_{2+})
			}{
				(\omega-\omega_{m1}+i\alpha_{0}+i\kappa_{m1})
				(\omega-\omega_{m2}+i\beta_{0}+i\kappa_{m2})
				+ \kappa_{m1}\kappa_{m2}
			}\left(1-e^{-\gamma_{+} t}\right).
		\end{equation}
	\end{widetext}
	
	This equation indicates that the envelope grows as $(1-e^{-\gamma_{+}t})$ with the power following its square, both exhibiting the same time scale $1/\gamma_{+}$. Fig.~\ref{fig3a3}(b) is the calculation of Eq.~(\ref{D2}) with the same previous parameters, the time scale $1/\gamma_{+}$ is demonstrated with open arrows. The same behavior occurs when probing at $\omega_{-}$ as shown in Fig.~\ref{fig3a3}(c) with the time scale $1/\gamma_{-}$ is demonstrated with open arrows. This confirms that $\gamma_{\pm}$ is associated with a time-scale for the transient buildup as well.
	
	\paragraph{Steady-state}
	In contrast to transient behavior, effective decay rates $\gamma_{\mathrm{eff},\pm}$ manifest themselves only in steady-state. The total output $|S_{tot}|^2=(|S_{1-}|^2+|S_{2-}|^2)/(|S_{1+}|^2+|S_{2+}|^2)=(|S_{1-}|^2+|S_{2-}|^2)/2|S_{1+}|^2)$ can be obtained by substituting Eq.~(\ref{D2}) into Eq.~(\ref{B2}) and (\ref{B3}). Due to large detuning, probing frequency $\omega_{+}\approx \omega_{+}'$ and neglecting the small mismatch between $\alpha_{0}$ and $\beta_{0}$, the total output can be expressed as
	\begin{equation}\label{C12}
		|S_{\mathrm{tot}}|^{2}
		=
		\left|
		\frac{
			\gamma_{\mathrm{eff,+}}
			\left(\sqrt{\Delta-4\Gamma}+i\gamma_{\mathrm{eff,-}}\right)
		}{
			\gamma_{+}
			\left(\sqrt{\Delta-4\Gamma}+i\gamma_{-}\right)
		}
		\right|^{2}.
	\end{equation}
	
	From this expression, although several parameters govern the steady-state output, under the CPA condition $\gamma_{\mathrm{eff},+}=0$ forces the total output to vanish, as demonstrated by the gray curve in Fig.~\ref{fig3a3}(d). A similar behavior is observed when probing at $\omega_{-}$, as indicated by the black dashed curve, where the steady-state output also vanishes.
	
	\section{Modes and anti-modes in the absorption}\label{abs}
	In prior work on indirectly coupled resonators in an open system \cite{lu2025temporal_indirect}, time-domain transient measurements revealed that the modes of such systems are encoded in their free temporal oscillations in the absence of continuous driving, the real and imaginary parts of the modes corresponding to the oscillation frequencies and decay rates, respectively. In the frequency domain, these same modes appear as the common poles of the S-parameters, the total output, and the absorption.
	
	For the same system under continuous excitation, when the ' extrinsic damping/coupling rates to the transmission channel are closely matched $\kappa_{m1}=\kappa_{m2}$, absorption maxima have been reported to indicate the frequency of the system mode \cite{wang_2024, lu2025temporal_indirect}. However, when this matching condition is not satisfied $\kappa_{m1} \neq \kappa_{m2}$, the absorption maxima are no longer guaranteed to track the mode frequencies, as observed in the present work by their alignment with the anti-mode frequencies. This motivates a careful examination of the underlying physical origin of absorption maxima.
	
	Here, we analyze the system's absorption under two identical inputs,
	\begin{equation}
		\begin{aligned}
			\mathrm{Abs}
			&= 1 - |S_{\mathrm{tot}}|^{2} \\
			&= 1 - \left|
			\frac{
				(\omega - \omega_{+}' + i\gamma_{\mathrm{eff},+})(\omega - \omega_{-}' + i\gamma_{\mathrm{eff},-})
			}{
				(\omega - \omega_{+} + i\gamma_{+})(\omega - \omega_{-} + i\gamma_{-})
			}
			\right|^{2},
		\end{aligned}
	\end{equation}
	where $\omega_{\pm}'$ and $\omega_{\pm}$ are the real parts, with
	$\gamma_{\mathrm{eff},\pm}$ and $\gamma_{\pm}$ the corresponding imaginary parts,
	of the zeros and poles, as discussed in Eqs.(\ref{E7})–(\ref{E9}).
	
	This equation shows that the absorption maxima are governed by the combined influence of both zeros and poles. From the linewidth perspective, the poles appear in the denominator and determine the primary spectral profile, while the zeros can introduce subtle secondary distortions; nevertheless, the apparent linewidth reflects the combined influence of $\gamma_{\pm}$ and $\gamma_{\mathrm{eff},\pm}$. From a frequency perspective, because the probe signal is purely real, the relative magnitudes of $\gamma_{\mathrm{eff},\pm}$ and $\gamma_{\pm}$ control the relative contributions of the zeros and poles to the absorption maxima. As a result, the absorption maxima are pulled toward the zero or pole with the dominant contribution.
	
	In most scenarios, the zeros lie closer to the real axis because $\gamma_{\pm}$ is typically larger than $\gamma_{\mathrm{eff},\pm}$ (see Fig.~\ref{fig7}(b) for an example); consequently, the maxima in the absorption spectrum primarily encode information about the zeros rather than the poles. Therefore, the absorption spectrum more readily reflects resonances of the anti-modes, not of the modes. In the extreme case where $\gamma_{\mathrm{eff},\pm}=0$, the zeros lie on the real axis, and the absorption maxima are purely determined by the zeros, as demonstrated in this work through CPA. However, when $\kappa_{m1} = \kappa_{m2}$, the real parts of the zeros and poles coincide, so the absorption maxima simultaneously indicate resonances of both modes and anti-modes \cite{lu2025temporal_indirect}.
	
	More generally, under arbitrary parameter choices, the interplay between zeros and poles can shift the absorption peak frequencies and affect the absorption linewidths. Consequently, the peak positions do not necessarily track mode or anti-mode resonances, and the measured linewidths do not simply correspond to $\gamma_{\pm}$ or $\gamma_{\mathrm{eff},\pm}$. In summary, the correspondence between absorption features and modes or anti-modes is strongly parameter dependent; therefore, absorption maxima do not provide a universal indicator of either modes or anti-modes.


\begin{thebibliography}{54}
\bibitem{dataavilibility}
		The data supporting the findings of this work will be made publicly available upon acceptance.
\end{thebibliography}
\end{document}